\documentclass[english,aps,prd,preprintnumbers,tightenlines,
showpacs,showkeys,floatfix,nofootinbib,superscriptaddress,fleqn]
{revtex4-1}
\usepackage{amsmath,amsfonts,amssymb,amscd,amsxtra,amsthm}
\usepackage{graphicx,slashed}
\usepackage{bm}
\usepackage{multirow}

\usepackage[normalem]{ulem} 
\usepackage[dvips]{color} 
\renewcommand\sout{\bgroup \color{red} \ULdepth=-.5ex \ULset}

\begin{document}   
\preprint{INHA-NTG-02/2014}
\title{Effects of $N(2000)\, 5/2^+$ , $N(2060)\,5/2^-$,
  $N(2120)\,3/2^-$, and $N(2190)\,7/2^-$ \\ on $K^*\Lambda$
  photoproduction}         
\author{Sang-Ho Kim}
\email[E-mail: ]{shkim@rcnp.osaka-u.ac.jp}
\affiliation{Research Center for Nuclear Physics (RCNP), Osaka
  567-0047, Japan}
\affiliation{Department of Physics, Inha University, Incheon 402-751, 
  Republic of Korea}
\author{Atsushi Hosaka}
\email[E-mail: ]{hosaka@rcnp.osaka-u.ac.jp}
\affiliation{Research Center for Nuclear Physics (RCNP), Osaka
  567-0047, Japan}   
\author{Hyun-Chul Kim}
\email[E-mail: ]{hchkim@inha.ac.kr}
\affiliation{Department of Physics, Inha University, Incheon 402-751,
  Republic of Korea}
\affiliation{School of Physics, Korea Institute for Advanced Study
  (KIAS), Seoul 130-722, Republic of Korea} 
\date{\today}
\begin{abstract}
We reinvestigate $K^*\Lambda(1116)$ photoproduction off the nucleon  
target, based on an effective Lagrangian approach. We include higher
nucleon resonances such as $N(2000)\, 5/2^+$, $N(2060)\,5/2^-$, 
$N(2120)\,3/2^-$, and $N(2190)\,7/2^-$, of which the data are taken 
from the 2012 edition of Review of Paritcle Physics, in addition to
the $t$-channel diagrams ($K$, $K^*$, and $\kappa$), the $s$-channel
nucleon, and $u$-channel hyperon($\Lambda$, $\Sigma$, and $\Sigma^*$) 
contributions. We find that the $N(2120)\,3/2^-$ and $N(2190)\,7/2^-$
resonances are essential in describing the new CLAS data for charged
$K^*$ photoproduction. On the other hand, they rarely affect for
neutral $K^*$ photoproduction.  
\end{abstract} 
\maketitle
\section{Introduction}
The CLAS collaboration at the Thomas Jefferson National Accelerator
Facility (TJNAF) has recently
reported the first high-statistics experimental data 
both for the total and differential cross sections
for the reaction $\gamma p \to K^{*+}\Lambda$~\cite{Wei:2013}. 
As compared to the previous preliminary data shown in the 
conference proceedings~\cite{Guo:2006kt}, the total cross section 
in the resonance region is significantly larger.
Though the original motivation of 
Ref.~\cite{Wei:2013} was to study the role of $\kappa(800)$ meson
involved in the $t$-channel process, the new CLAS data near the
threshold gives us a clue in understanding the role of higher nucleon
($N^*$) resonances. In a previous work~\cite{Kim:2011rm}, it was
found that the $N^*$ resonances indeed played an important role in
describing the experimental data near the threshold region. However,
the new CLAS data indicates that there are still missing part in the
previous analysis. 
As discussed in Ref.~\cite{Wei:2013} in detail, all
theoretical results~\cite{Oh:2006hm,Oh:2006in,Ozaki:2009wp,Kim:2011rm} 
look different from the CLAS data. In this respect, it is of great
importance to reinvestigate the production mechanism of 
$K^{*+}\Lambda$ photoproduction. In Ref.~\cite{Kim:2011rm}, it was
pointed out that certain $N^*$ resonances are essential in describing 
the former experimental data near the threshold region. In particular, 
$D_{13}(2080)$ was shown to be crucial in explaining the
enhancement of the near-threshold production rate. 

In the meanwhile, the data for the $N^*$ resonances in the
2012 edition of Review of Particle Physics~\cite{Beringer:2013ab} were
much changed from those in the 2010 edition~\cite{Nakamura:2010zzi}.  
This revision is mainly due to a new multi-channel partial wave
analysis~\cite{Anisovich:2013cd}. So far the evidence and
properties of $N^*$ resonances were determined by the partial wave
analyses of $\pi N$ scattering data~\cite{Hohler:1979yr} but they are
still far from complete understanding. Anisovich et al. performed
a multichannel partial wave analysis taking both of the $\pi N$ and
various photoproduction data~\cite{Anisovich:2013cd}. Based on this
analysis, a few new $N^*$ resonances were included and some were
rearranged in the $N^*$ spectrum~\cite{Beringer:2013ab}. In
particular, four new $N^*$ resonances were classified below 1.9 GeV:
$N(1860)5/2^+$, $N(1875)3/2^-$, $N(1880)1/2^+$, and
$N(1895)1/2^-$~\cite{Anisovich:2011sv,Anisovich:2011su}. Some of the
$N^*$ resonances above the $K^*\Lambda$ threshold were either newly
found or rearranged. For example, the mass of the $D_{15}(2200)$ was
moved down to $N(2060)\,5/2^-$ with its photon decay amplitudes
added. As for the $N(2190)\,7/2^-$, its photon decay amplitudes were
renewed. A noticeable thing is that the $D_{13}(2080)$
has disappeared in the PDG 2012 edition. Instead, two new resonances
with $J^P=3/2^-$ are included: $N(1875)3/2^-$ and $N(2120)5/2^-$. The
old $D_{13}(2080)$ seems to correspond to  $N(1875)3/2^-$ below the 
$K^*\Lambda$ threshold, though the new data of the photon decay
helicity amplitudes~\cite{Anisovich:2013cd,Beringer:2013ab} are very
different from the old ones~\cite{Awaji:1981zj,
  Fujii:1981kx,Nakamura:2010zzi}. If one takes this situation
seriously, we have to reanalyze the production mechanism of the
$\gamma N\to K^* \Lambda$ with the new $N^*$ data employed. 

In the present work, we reexamine $K^*\Lambda(1116)$
photoproduction off the nucleon, considering some of the $N^*$  
resonances of the PDG 2012 edition above the threshold. We will take
$N(2000)5/2^+$, $N(2060)5/2^-$, $N(2120)3/2^-$, and $N(2190)7/2^-$
into account. The last one was omitted in the previous
analysis~\cite{Kim:2011rm} because of the complexity  
due to its higher spin. The $P_{11}(2100)$ is not included here
because of the lack of information. To reduce the
ambiguity in determining the coupling constants, we use the
experimental data when they are available. 
As we will show later, the results of the total cross section for the
$\gamma p\to K^{*+} \Lambda$ reaction are in  remarkable agreement 
with the new CLAS data~\cite{Wei:2013}. Its differential cross
sections are also well reproduced, compared to those from previous 
works~\cite{Oh:2006hm,Oh:2006in, Ozaki:2009wp,Kim:2011rm}.  We
predict the total and differential cross sections of the neutral
process $\gamma n\to K^{*0} \Lambda$. Anticipating the results from
future experiments, we compute the beam, recoil and target
asymmetries of $\gamma N\to K^*\Lambda$ reactions. In addition, we
derive some of the double polarization observables.  

We sketch the present paper as follows: In Sec.~II, we briefly explain 
the general framework.  The effective Lagrangians required for
$K^*\Lambda$ photoproduction are presented explicitly. We also
describe how to fix the coupling constants and the cut-off masses.
In Sec.~III, the results of the cross sections are compared with the
experimental data for the $\gamma p \to K^*\Lambda$ reaction. 
We also show the predictions of the polarization observables and
discuss them. Section~IV is devoted to summary and draws conclusions. 

\section{formalism}
\begin{figure}[ht]
\includegraphics[scale=0.5]{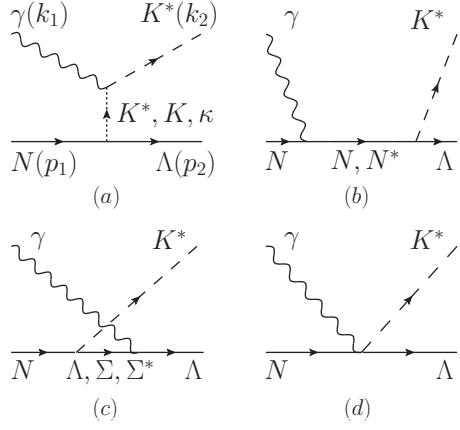}
\caption{The tree-level Feynman diagrams for the 
         $\gamma N\to K^*\Lambda$ reaction.}        
\label{FIG1}
\end{figure}
In this section, we briefly explain the general formalism of an
effective Lagrangian approach. We refer to Ref.~\cite{Kim:2011rm} for
more details. The tree-level Feynman diagrams relevant to the $\gamma
N\to K^*\Lambda$ reaction is shown in Fig.~\ref{FIG1}. $k_1$ and 
$p_1$ denote respectively the momenta of incoming photon and nucleon,
while $k_2$ and $p_2$ represent those of the outgoing $K^*$ and
$\Lambda$, respectively. Diagram (a) stands for the $t$-channel
processes including $K^*$, $K$, and $\kappa$ exchange, diagram (b)
shows the $s$-channel processes containing the nucleon and $N^*$
resonances, diagram (c) corresponds to the $u$-channel ones with 
$\Lambda$, $\Sigma$, and $\Sigma^*$ exchanges, and diagram (d) is the 
contact term required to preserve gauge invariance. 

The basic form of the effective Lagrangians are already given in
previouse works. The photon-meson interactions are described by the
following effective Lagrangians 
\begin{eqnarray}
\label{eq:LAG1}
\mathcal L_{\gamma K^* K^*} &=& 
-ie_{K^*} A^\mu \left( K^{*\nu} K_{\mu\nu}^{*\dagger} -
                K_{\mu\nu}^* K^{*\dagger\nu} \right),
\nonumber \\ 
\mathcal L_{\gamma K^* K} &=&
g_{\gamma K^* K} \varepsilon^{\mu\nu\alpha\beta} 
\left( \partial_\mu A_\nu \right) 
\left( \partial_\alpha K_\beta^* \right) 
\bar K + \mathrm{h.c.},
\nonumber \\
\mathcal{L}_{\gamma K^* \kappa} &=&
g_{\gamma K^* \kappa} A^{\mu\nu} \bar \kappa 
K_{\mu\nu }^* + \mathrm{h.c.},
\end{eqnarray}
where $A_\mu$, $K^*_\mu$, $K$, and $\kappa$ denote the photon, the
$K^*(892,1^-)$ vector meson, the $K(495,0^-)$ pseudoscalar meson, and
the $\kappa(800,0^+)$ scalar meson, respectively. The $K_{\mu\nu}^*$
represents the field-strength tensor for the $K^*$ vector meson
defined as $K^*_{\mu\nu}=\partial_\mu K^*_\nu-\partial_\nu K^*_\mu$.   
The electric charge of the $K^*$ vector meson is given as
$e_{K^*}^{}$. We take the values of $g_{\gamma K^*K}$ from the
experimental data from the PDG~\cite{Beringer:2013ab},
which lead to $g^\mathrm{charged}_{\gamma K^*K}=0.254$~GeV$^{-1}$ and   
$g^\mathrm{neutral}_{\gamma K^*K}=-0.388$~GeV$^{-1}$. On the other
hand, we utilize the vector-meson dominance~\cite{BHS02} to find the 
values of $g_{\gamma K^* \kappa}^{}$: $g^\mathrm{charged}_{\gamma
  K^*\kappa}=0.12\,e$~GeV$^{-1}$ and $g^\mathrm{neutral}_{\gamma
  K^*\kappa}=-0.24\,e $~GeV$^{-1}$ with the unit electric charge
  $e$.  

The effective Lagrangians for the electromagnetic (EM) interactions
for the baryons are given as  
\begin{eqnarray}
\label{eq:LAG2}
\mathcal {L}_{\gamma NN}&=&
- \bar N \left[ e_N \gamma_\mu - \frac{e\kappa_N}{2M_N}
\sigma_{\mu\nu}\partial^\nu \right] A^\mu N,
\nonumber \\
\mathcal {L}_{\gamma\Lambda\Lambda}&=&
\frac{e \kappa_\Lambda}{2M_N} \bar \Lambda
\sigma_{\mu\nu} \partial^\nu A^\mu \Lambda,
\nonumber \\
\mathcal {L}_{\gamma\Lambda\Sigma}&=&
\frac{e \mu_{\Sigma\Lambda}}{2M_N} \bar \Sigma \sigma_{\mu\nu}
\partial^\nu A^\mu \Lambda + \mathrm{h.c.},
\nonumber \\
\mathcal {L}_{\gamma\Lambda\Sigma^*}&=&
-\frac{ie}{2M_N} 
\left[ g^V_{\gamma\Lambda\Sigma^*} \bar\Lambda \gamma_\nu -
\frac{ig^T_{\gamma\Lambda\Sigma^*}}{2M_N} \partial_\nu
\bar \Lambda \right]
\gamma_5\Sigma_\mu^*F^{\mu\nu} + \mathrm{h.c.},
\end{eqnarray}
where $N$, $\Lambda$, $\Sigma$, and $\Sigma^*$ stand for the nucleon, 
$\Lambda(1116)$, $\Sigma(1192)$, and $\Sigma^*(1385,3/2^+)$ hyperon
fields, respectively. $M_h$ denotes generically the mass of hadron
$h$. The baryon fields with spin $s = 3/2$ are described by the 
Rarita-Schwinger field~\cite{Rarita:1941mf,Read:1973ye}. Here,
$\kappa_B^{}$ is the anomalous magnetic moment of a baryon $B$ and
$\mu_{\Lambda\Sigma}$ designates the transition magnetic moment
between the $\Lambda(1116)$ and the $\Sigma(1192)$. Note that the EM
couplings for the spin-3/2 hyperon $\Sigma^*$ are related to the
well-known magnetic dipole ($M1$) and electric quadrupole ($E2$)
moments. These coupling constants are determined by the experimental
data of the radiative decay width $\Gamma_{\Sigma^*\to
  \gamma\Lambda}$~\cite{Beringer:2013ab}, which 
leads to $(g^V_{\gamma\Lambda\Sigma^*}, g^T_{\gamma\Lambda\Sigma^*}) =
(3.78,3.18)$.   

The effective Lagrangians for the meson-baryon interactions are
\begin{eqnarray}
\label{eq:LAG3}
\mathcal L_{KN\Lambda} &=&
-ig_{KN\Lambda} \bar N \gamma_5 \Lambda K + \mathrm{h.c.},       
\nonumber \\
\mathcal{L}_{\kappa N\Lambda} &=&
-g_{\kappa N\Lambda} \bar N \Lambda \kappa + \mathrm{h.c.},     
\nonumber \\
\mathcal L_{K^* NY} &=& 
-g_{K^* NY} \bar N \left[ \gamma_\mu Y - 
\frac{\kappa_{K^* NY}}{2M_N} \sigma_{\mu\nu} Y \partial^\nu \right] 
K^{*\mu} + \mathrm{h.c.}, 
\nonumber \\
\mathcal L_{K^* N\Sigma^*}&=& 
-\frac{if^{(1)}_{K^* N\Sigma^*}}{2M_{K^*}} \bar K^*_{\mu\nu} 
\bar \Sigma^{*\mu} \gamma^\nu \gamma_5 N -
\frac{f^{(2)}_{K^* N\Sigma^*}}{4M_{K^*}^2} \bar K^*_{\mu\nu}
\bar \Sigma^{*\mu} \gamma_5 \partial^\nu N +
\frac{f^{(3)}_{K^* N\Sigma^*}}{4M_{K^*}^2} \partial^\nu 
\bar K^*_{\mu\nu} \bar \Sigma^{*\mu} \gamma_5 N
+ \mathrm{h.c.}     ,
\end{eqnarray}
where $\Sigma = \bm{\tau} \cdot \bm{\Sigma}$ and $\Sigma^*_\mu =
\bm{\tau} \cdot \bm{\Sigma}^*_\mu$. The strong coupling constants are
mainly determined by the flavor SU(3) symmetry and hypeon-nucleon
potential models (for example, the Nijmegen potential~\cite{RSY99}). 
Considering the Lorentz structure for the vector-meson coupling to the
$\Sigma^*$, we can write the interaction Lagrangian 
in terms of the three form factors, which are similar to the case of  
$\mathcal{L}_{\gamma\Lambda\Sigma^*}$. From the flavor SU(3)
symmetry, the value for $f^{(1)}_{K^* N\Sigma^*}$ can be
estimated. Because of the lack of experimental and theoretical 
information on $f^{(2,3)}_{K^* N\Sigma^*}$, we ignore these
terms in the present work, which is plausible, since these two
coupling constants are smaller than $f_{K^*N\Sigma^*}^{(1)}$.
Finally, the contact term should be included only for the charged
$K^*$ production to preserve the U(1) gauge invariance. The
corresponding Lagrangian is written as  
\begin{equation}
\label{eq:LAG4}
\mathcal L_{\gamma K^* N\Lambda}=
-\frac{ie_{K^*} g_{K^*N\Lambda} \kappa_{K^* N\Lambda}}{2M_N}
\bar \Lambda \sigma^{\mu\nu} A_\nu K^{*}_\mu N + \mathrm{h.c.}.
\end{equation} 
As for the details of the relevant coupling constants and other
parameters, we refer to Ref.~\cite{Kim:2011rm}.

In addition to the effective Lagrangians for the basic processes
discussed above, we now consider those for the $N^*$ resonances.
The EM Lagrangians for the $N^*$ resonances 
from spin $1/2$ to $7/2$ are given as
\begin{eqnarray}
\label{eq:LAGR1}
\mathcal{L}_{\gamma  N N^*} \left(\frac{1}{2} \right)^\pm  &=& 
\frac{eh_1}{2M_N} \bar N \Gamma^{(\mp)}
\sigma_{\mu\nu} \partial^\nu A^\mu N^* + 
\mathrm{h.c.},                       
\cr
\mathcal{L}_{\gamma N N^*} \left(\frac{3}{2} \right)^\pm &=& 
-ie \left[ \frac{h_1}{2M_N} \bar N \Gamma_\nu^{(\pm)} - 
           \frac{ih_2}{(2M_N)^2} \partial_\nu \bar N
           \Gamma^{(\pm)} \right] F^{\mu\nu} N^*_\mu + 
\mathrm{h.c.},                       
\cr
\mathcal{L}_{\gamma N N^*} \left(\frac{5}{2} \right)^\pm &=& 
e \left[ \frac{h_{1}}{(2M_N)^2} \bar N \Gamma_\nu^{(\mp)} -
         \frac{ih_{2}}{(2M_N)^3} \partial_\nu \bar N
         \Gamma^{(\mp)} \right] \partial^\alpha F^{\mu\nu} 
         N^*_{\mu\alpha} + 
\mathrm{h.c.},
\cr
\mathcal{L}_{\gamma N N^*} \left(\frac{7}{2} \right)^\pm &=& 
ie \left[ \frac{h_{1}}{(2M_N)^3} \bar N \Gamma_\nu^{(\pm)} -
          \frac{ih_{2}}{(2M_N)^4} \partial_\nu \bar N
          \Gamma^{(\pm)} \right] \partial^\alpha \partial^\beta 
          F^{\mu\nu} N^*_{\mu\alpha\beta} +
\mathrm{h.c.},   
\end{eqnarray}
where $N^*$ denotes the corresponding nucleon resonance field. 
The $\Gamma^{(\pm)}$ and the $\Gamma_\mu^{(\pm)}$ are defined, 
respectively, as  
\begin{equation}
\label{eq:GAMMASPEM}
\Gamma^{(\pm)} = \left(
\begin{array}{c} 
\gamma_5 \\ \mathbf{1}
\end{array} \right),
\qquad
\Gamma_\mu^{(\pm)} = \left(
\begin{array}{c}
\gamma_\mu \gamma_5 \\ \gamma_\mu 
\end{array} \right).
\end{equation}
The effective Lagrangians for the strong vertices including the
$N^*$ resonances are expressed as 
\begin{eqnarray}
\label{eq:LAGR2}
\mathcal{L}_{ K^* \Lambda N^*} \left(\frac{1}{2} \right)^\pm
&=&-\frac{1}{2M_N} \bar N^* \left[
g_1 \left( \pm \frac{\Gamma_\mu^{(\mp)} \Lambda \partial^2}{M_R \mp M_N}
-i \Gamma^{(\mp)} \partial_\mu \right)
-g_2 \Gamma^{(\mp)} \sigma_{\mu\nu} \Lambda \partial^\nu 
\right] K^{*\mu} + 
\mathrm{h.c.},
\cr
\mathcal{L}_{ K^* \Lambda N^*} \left( \frac{3}{2} \right)^\pm                
&=& i \bar N^*_\mu \left[ 
\frac{g_1}{2M_N} \Lambda \Gamma_\nu^{(\pm)} \mp 
\frac{ig_2}{(2M_N)^2} \partial_\nu \Lambda \Gamma^{(\pm)} \pm 
\frac{ig_3}{(2M_N)^2} \Lambda \Gamma^{(\pm)} \partial_\nu 
\right] K^{*\mu\nu} + 
\mathrm{h.c.},                                       
\cr
\mathcal{L}_{ K^* \Lambda N^*} \left(\frac{5}{2} \right)^\pm                   
&=& \bar N^*_{\mu\alpha} \left[ 
\frac{g_1}{(2M_N)^2} \Lambda \Gamma_\nu^{(\mp)} \pm 
\frac{ig_2}{(2M_N)^3} \partial_\nu \Lambda \Gamma^{(\mp)} \mp 
\frac{ig_3}{(2M_N)^3} \Lambda \Gamma^{(\mp)} \partial_\nu 
\right] \partial^\alpha K^{*\mu\nu} + 
\mathrm{h.c.},           
\cr
\mathcal{L}_{ K^* \Lambda N^*} \left(\frac{7}{2} \right)^\pm
&=& -i \bar N^*_{\mu\alpha\beta} \left[ 
\frac{g_1}{(2M_N)^3} \Lambda \Gamma_\nu^{(\pm)} \mp 
\frac{ig_2}{(2M_N)^4} \partial_\nu \Lambda \Gamma^{(\pm)} \pm 
\frac{ig_3}{(2M_N)^4} \Lambda \Gamma^{(\pm)} \partial_\nu 
\right] \partial^\alpha \partial^\beta K^{*\mu\nu} + 
\mathrm{h.c.}. 
\end{eqnarray}
The $N^*$ resonance field for a spin of 3/2 is treated 
as the Rarita-Schwinger field~\cite{Rarita:1941mf,Read:1973ye}, so
that the corresponding propagator with momentum $p$ and
mass $M$ is written as 
\begin{eqnarray}
\label{eq:RSSP}
&&\Delta_{\alpha\beta}(p,M)=\frac{i(\rlap{/}{p}+M)}{p^2-M^2}
\left[ - g_{\alpha\beta} +\frac{1}{3}\gamma_\alpha \gamma_\beta  +
\frac{1}{3M} (\gamma_\alpha p_\beta - \gamma_\beta p_\alpha )
 +\frac{2}{3M^2}p_\alpha p_\beta \right].
\end{eqnarray}
The propagators of the $N^*$ resonance fields for spins of 5/2 and 7/2 
are expressed~\cite{Chang:1967ab, Rushbrooke:1966cd, Behrends:1957ef,
  Oh:2012gh} as
\begin{eqnarray}
\label{eq:RSSP2}
&&\Delta_{{\alpha_1}{\alpha_2};{\beta_1}{\beta_2}}(p,M)=
\frac{i(\rlap{/}{p}+M)}{p^2-M^2}                                     \cr
&&\times
\left[\frac{1}{2}( \bar g_{{\alpha_1}{\beta_1}} \bar g_{{\alpha_2}{\beta_2}} 
+\bar g_{{\alpha_1}{\beta_2}} \bar g_{{\alpha_2}{\beta_1}})
-\frac{1}{5} \bar g_{{\alpha_1}{\alpha_2}} \bar g_{{\beta_1}{\beta_2}} 
-\frac{1}{10}( \bar \gamma_{\alpha_1} \bar \gamma_{\beta_1} \bar
g_{{\alpha_2}{\beta_2}}  
+\bar \gamma_{\alpha_1} \bar \gamma_{\beta_2} \bar g_{{\alpha_2}{\beta_1}}
+\bar \gamma_{\alpha_2} \bar \gamma_{\beta_1} \bar g_{{\alpha_1}{\beta_2}} 
+\bar \gamma_{\alpha_2} \bar \gamma_{\beta_2} \bar g_{{\alpha_1}{\beta_1}} ) 
\right],                                                            \cr
&&\Delta_{{\alpha_1}{\alpha_2}{\alpha_3};{\beta_1}{\beta_2}{\beta_3}}
(p,M)=\frac{i(\rlap{/}{p}+M)}{{p^2-M^2}}                             \cr
&&\times
\frac{1}{36}\sum_{P(\alpha),P(\beta)}
\left[-\bar g_{{\alpha_1}{\beta_1}} \bar g_{{\alpha_2}{\beta_2}} 
\bar g_{{\alpha_3}{\beta_3}} +\frac{3}{7}\bar g_{{\alpha_1}{\beta_1}}
\bar g_{{\alpha_2}{\alpha_3}} \bar g_{{\beta_2}{\beta_3}}
 +\frac{3}{7} \bar \gamma_{\alpha_1} \bar \gamma_{\beta_1} 
 \bar g_{{\alpha_2}{\beta_2}} \bar g_{{\alpha_3}{\beta_3}}
 -\frac{3}{35} \bar \gamma_{\alpha_1} \bar \gamma_{\beta_1} 
 \bar g_{{\alpha_2}{\alpha_3}} \bar g_{{\beta_2}{\beta_3}} \right],
\end{eqnarray}
where the sum is over all permutations of $\alpha$'s and $\beta$'s, and
\begin{equation}
\bar{g}_{\alpha\beta} = g_{\alpha\beta} - \frac{p_\alpha p_\beta}{M^2}, 
\,\,\,\,
\bar{\gamma}_\alpha = \gamma_\alpha - \frac{p_\alpha}{M^2}\rlap{/}{p}.
\end{equation}
Here, the mass of the $N^*$ resonance in the $N^*$ propagator is   
replaced as $M \rightarrow M-i\Gamma/2$ with its decay width
$\Gamma$. In gegeral, off-shell parameters may appear in resonance
propagators and vertices. However, such off-shell effects are not
significant because resonances come into play near the on-mass shell
region~\cite{Nam:2005op}, which has been verified numerically.

We employ the data for the $N^*$ resonances taken from the PDG 2012
edition~\cite{Beringer:2013ab}, as mentioned in detail in Introduction. 
We consider in this work $N(2000)\,5/2^+$, $N(2060)\,5/2^-$, 
$N(2120)\,3/2^-$, and $N(2190)\,7/2^-$ near the threshold region. 
The values of the masses and decay widths are taken from the 
Breit-Wigner values~\cite{Anisovich:2013cd,Anisovich:2013ef}. 
The transition magnetic moments $h_1$ and $h_2$ given in 
Eq.~(\ref{eq:LAGR1}) are determined by the Breit-Wigner helicity 
amplitudes taken
from Refs.~\cite{Anisovich:2013cd,Anisovich:2013ef} or 
by the predictions from the relativistic quark 
model~\cite{Capstick:1992uc}:   
the parameters for $N(2000)\,5/2^+$, $N(2060)\,5/2^-$, and 
$N(2120)\,3/2^-$ are taken from
Refs.~\cite{Anisovich:2013cd,Anisovich:2013ef}, whereas   
those for $N(2190)\,7/2^-$ are determined by using the results in
Ref.~\cite{Capstick:1992uc}. 

The strong coupling constants in Eq.~(\ref{eq:LAGR2}), $g_i$, are 
found by the following relation,
\begin{equation}
\label{eq:GGG}
\Gamma_{N^*\to K^*\Lambda}=\sum_{l,s}|G(l,s)|^2,
\end{equation}
where the explicit form of the decay amplitudes $G(l,s)$ is 
given in Ref.~\cite{Capstick:1998uh}. Here, we take into account the
lowest partial-wave contribution for  
$G(l,s)$ and therefore only the lowest multipole, i.e., the first term
of Eq.~(\ref{eq:LAGR2}), is considered as in Ref.~\cite{Kim:2011rm}. 
This assumption is reasonable, as will be shown in the next
section. The signs of these strong coupling constants are determined
phenomenologically. Because of lack of information, we also assume
that $N(2000)\,5/2^+$, $N(2060)\,5/2^-$, $N(2120)\,3/2^-$, and
$N(2190)7/2^-$ may correspond respectively to $F_{15}(2000)$,
$D_{15}(2200)$, $D_{13}(2080)$, and $G_{17}(2190)$ in the PDG 2010 
edition~\cite{Nakamura:2010zzi}. However, as will be discussed in the
next section, the $N(2120)\,3/2^-$ turns out to be distinguished from
the old $D_{13}(2080)$ that played an important role in the previous
work~\cite{Kim:2011rm}. In fact, the $D_{13}(2080)$ more or less
corresponds to the lower-lying 3-star $N^*$ resonance
$N(1875)\,3/2^-$. Thus, we have to fit the parameters of the
$N(2120)\,3/2^-$ to the experimental data. Table~\ref{TABLE1} list the
relevant parameters for the $N^*$ resonances used in this work. 
\begin{table}[hptb]
\begin{tabular}{cccc|cccc|ccc} \hline\hline
&$\hspace{0.2cm}\mathrm{PDG}\hspace{0.2cm}$
&$\mathrm{M_{BW}}$
&$\hspace{0.3cm}\Gamma_{BW}\hspace{0.3cm}$
&$\hspace{0.7cm}A_1\hspace{0.7cm}$
&$\hspace{0.7cm}A_3\hspace{0.7cm}$ 
&$\hspace{1.1cm}h_1\hspace{1.1cm}$
&$\hspace{1.1cm}h_2\hspace{1.1cm}$
&$\hspace{0.4cm}G(l,s)\hspace{0.4cm}$
&$\hspace{0.2cm}g_1\hspace{0.2cm}$
&$\hspace{0.2cm}g_1(\mathrm{final})\hspace{0.2cm}$ \\
\hline
&$N(2000)\,5/2^+ $&$2090$ & $460$&$+32\,\,(-18) $&$+48\,\,(-35)$
&$+0.114(-0.395)$&$+1.22(-0.500)$&$+0.3$&$+1.37$&$+1.37$\\
&$N(2060)\,5/2^-$&$2060$ & $375$&$+67\,\,(+25)$&$+55\,\,(-37)$ 
&$-2.45(+0.027)$&$-3.81(-2.85)$ &$+0.2$&$+5.42$&$+5.42$\\ 
&$N(2120)\,3/2^-$&$2150$ & $330$&$+130\,\,(+110)$&$+150\,\,(+40)$
&$-0.827(-1.66)$&$+2.14(+2.31)$ &$+3.8$&$+1.29$&$+0.30$\\
&$N(2190)\,7/2^-$&$2180$ & $335$&$-34\,\,(+10)$&$+28\,\,(-14)$ 
&$+7.87(-2.94)$ &$-7.36(+2.49)$ &$+2.5$&$-44.3$&$-44.3$\\ 
\hline\hline
\end{tabular}
\caption{The masses, the decay widths, and 
the relevant parameters for the $N^*$ resonances. The helicity
amplitudes $A_{1,3}$ [$10^{-3}\mathrm{GeV}^{-\frac{1}{2}}$] are
obtained from Refs.~\cite{Anisovich:2013cd,Anisovich:2013ef, 
Capstick:1992uc} and the decay amplitudes $G(l,s)$ 
[$\mathrm{GeV}^{\frac{1}{2}}$] are estimated from Ref.
~\cite{Capstick:1998uh}. Those in the parentheses correspond 
to the neutron resonances.} 
\label{TABLE1}
\end{table}

Using the Lagrangians for the various vertices, we can obtain the
scattering amplitudes of $t$-, $s$- and $u$-channels, and contact
terms, where the $s$-channel includes $N^*$ resonances. Furthermore,
for the $K^*$, $\kappa$, and $\Sigma^*$ exchanges, we take into
account the decay widths in their propagators, 50.8, 550, and 36 Mev,
respectively~\cite{Beringer:2013ab}. 

In an effective Lagrangian approach, it is essential to consider a
form factor at each vertex, since it parameterizes the structure of
the hadron. However, it is in fact rather difficult to handle the form
factors at an EM vertex, since it is well known that it breaks the
gauge invariance due to its nonlocality~\cite{peierls}.   
To circumvent this problem, we follow a prescription explained in 
Refs.~\cite{Haberzettl:1998eq,Davidson:2001rk,Haberzettl:2006bn}. 
Though it is phenomenological, it provides a convenient way of handling 
the form factors for an EM vertex. The form factors for off-shell mesons
and baryons are given respectively as    
\begin{equation}
\label{eq:FF}
F_\Phi = \frac{\Lambda^2_\Phi - M^2_\Phi}{\Lambda^2_\Phi-p^2},
\quad
F_B = \frac{\Lambda^4_B}{\Lambda^4_B+(p^2-M^2_B)^2},
\end{equation}
where $M_{(\Phi,B)}$, and $p$ denote the the mass and the momentum of the 
off-shell hadron, respectively. In order to preserve gauge
invariance for the charged $K^*$ production, we consider a common form
factor for $K^*$ and $N$ exchanges as  
\begin{equation}
F_\mathrm{com}
= F_{K^*}F_N - F_{K^*}-F_N.
\end{equation}
The neutral $K^*$ production does not require this. The cut-off
parameters are determined phenomenologically. However, to reduce
theoretical ambiguities due to the wide range of the cut-off 
values, we limit their values around $1$ GeV.
\section{Results and Discussion}
Before we start the detailed discussions for the present results in
comparison with the new data, we would like to briefly summarize the
current and past situations of the experimental and theoretical
studies for $\gamma p \to K^{*+} \Lambda$ reaction.
Before the new CLAS data was announced in Ref.~\cite{Wei:2013}, there 
were two preliminary data that were already reported in
Refs.~\cite{Guo:2006kt,Hicks:2011ab}. 
In Ref.~\cite{Guo:2006kt}, only the preliminary total cross sections
were shown, while in Ref.~\cite{Hicks:2011ab} only the differential
cross sections were presented. Using those preliminary data, the
previous theoretical studies were done~\cite{Kim:2011rm,Oh:2006hm}.  
Now in the new CLAS data, both the total and differential cross
sections are given.

\begin{figure}[h]
\vspace{1.5em}
\begin{tabular}{cc} 
\includegraphics[scale=0.5]{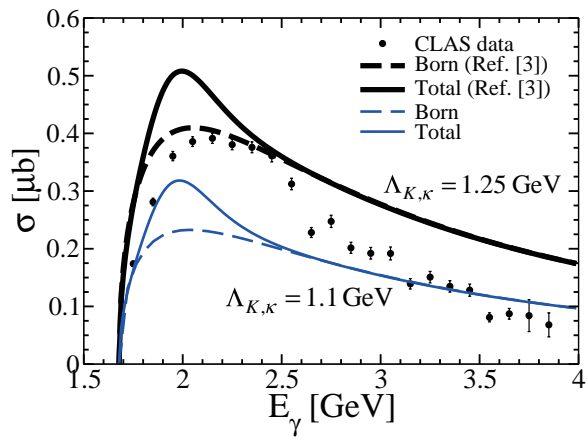}
\end{tabular}
\caption{(Color online)Total cross sections when using $\Lambda_{ K,\kappa}
=1.25$ GeV (thicker curves)  and 1.1 GeV (thinner ones). 
The dashed curves are the results only with the Born terms, and the 
solid ones for the inclusion of resonances of Ref.~\cite{Kim:2011rm}.} 
\label{FIG2}
\end{figure}

One of the differences between Ref.~\cite{Kim:2011rm} and 
Ref.~\cite{Oh:2006hm} is the choice of the cut-off values for the $K$ 
and $\kappa$ exchanges in the $t$-channel, which are 
$\Lambda_{K,\kappa}$=1.25 GeV and $\Lambda_{K,\kappa}$=1.1 GeV, respectively, 
to reproduce the used experimental data.
Another difference is that in Ref.~\cite{Kim:2011rm}, resonances are 
included, while in Ref.~\cite{Oh:2006hm} they are not.
Now using the new data, it turns out that the cutoff $\Lambda_{K,\kappa}$
should be taken at around 1.1 GeV. 
In Fig.~\ref{FIG2}, we show
total cross sections when using $\Lambda_{ K,\kappa} =1.25$ GeV
(thicker curves) and 1.1 GeV (thinner ones).

As we can see from this figure, the result of $\Lambda_{K,\kappa} = 1.25$ 
GeV overestimates the total cross section. 
The one of $\Lambda_{K,\kappa} = 1.1$ GeV agrees better especially at 
higher energies.   
The discrepancy near the threshold region is improved by the nucleon 
resonances but not enough to describe the experimental data, which is 
the issue of the present paper.
\begin{figure}[h]
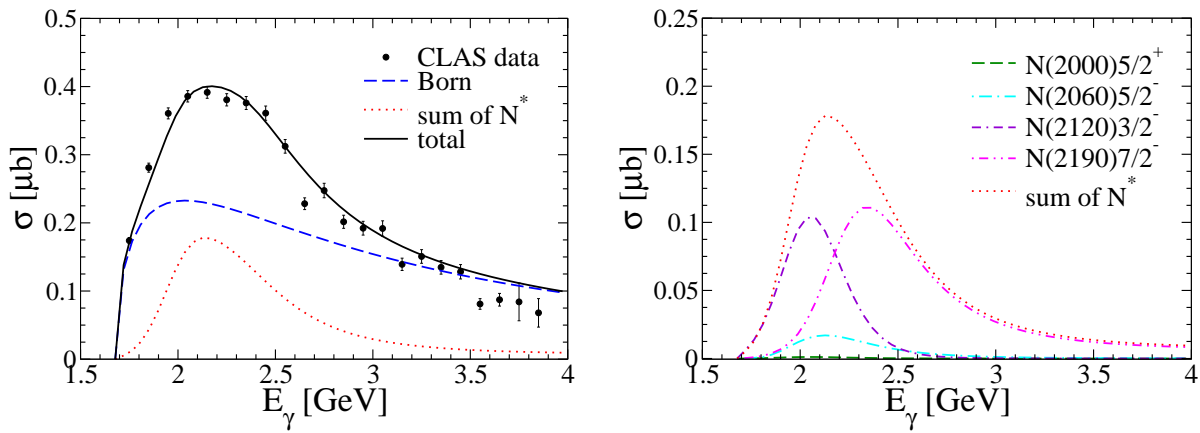

\vspace{1.5em}
\begin{tabular}{cc} 
\includegraphics[scale=0.3]{FIG3a.eps} \;\;\;\;
\includegraphics[scale=0.3]{FIG3b.eps}
\end{tabular}
\caption{(Color online) The results of the total
  cross sections for the  $\gamma p \to K^{*+}\Lambda$ reaction in the
  left panel. The dashed curve includes the Born diagrams presented in 
  Fig.~\ref{FIG1} without the $N^*$ resonances. The dotted curve shows 
  the contribution of the $N^*$ resonances. The solid curve draws the 
  total contribution of all diagrams. The black circles denote the new 
  CLAS data~\cite{Wei:2013}. The right panel illustrates each 
  contribution of the $N^*$ resonances.}         
\label{FIG3}
\end{figure}

Now in Fig.~\ref{FIG3}, we show the new results for the total 
cross section with various contributions. 
In the left panel, the total cross section with the background
contribution and with the total contributions of all the $N^*$ 
resonances are shown.
The dashed curve includes the Born diagrams presented in 
Fig.~\ref{FIG1} without the $N^*$ resonances.
The result more or less corresponds to that of Ref.~\cite{Oh:2006hm}.
The dotted one depicts the contribution of the $N^*$ resonances.
The cut-off parameters are used as
$\Lambda_{K^*,N,\Lambda,\Sigma,\Sigma^*}=0.9$ GeV,
$\Lambda_{K,\kappa}=1.1$ GeV, and $\Lambda_R=1.0$ GeV. 
With the inclusion of the $N^*$ resonances, the present theoretical 
result is drawn as the solid curve, which describes very well the 
experimental data.

Let us discuss the details of resonance contributions as shown 
in the right panel of Fig.~\ref{FIG3}.
The contribution from $N(2000)\,5/2^+$ turns out to be almost negligible.
The $N(2060)\,5/2^-$ makes a small contribution to the total cross
section. Concerning $N(2120)\,3/2^-$, we first assume that it
corresponds to the old $D_{13}(2080)$, since their masses are similar
each other with the same spin quantum numbers.
As done in Ref.~\cite{Kim:2011rm}, we have computed the effect of
$N(2120)\,3/2^-$ but it turns out to be overestimated in comparison 
with the experimental data. In fact, it has yielded approximately
$\sim 1.9\, \mu b$ for the total cross section. 
Thus, we determine the strong coupling constant of $N(2120)\,3/2^-$ by
fitting it to the data for the total cross section. As a result, the
coupling constant $g_1$ is changed from $+1.29$ to $+0.30$, as shown
in Table~\ref{TABLE1}. Bearing in mind this fact, we show in the right
panel of Fig.~\ref{FIG3} that the $N(2120)\,3/2^-$ makes an important
contribution to the total cross section with a peak around 2 GeV. The
$N(2190)7/2^-$ also turns out to be as equally important as
$N(2120)\,3/2^-$. In particular, it governs the dependence of the
total cross section on the photon energy $E_\gamma$ in higher
$E_\gamma$ regions. With these two $N^*$ resonances taken into
account, the experimental data for the total cross section is well
reproduced. 
\begin{figure}[ht]
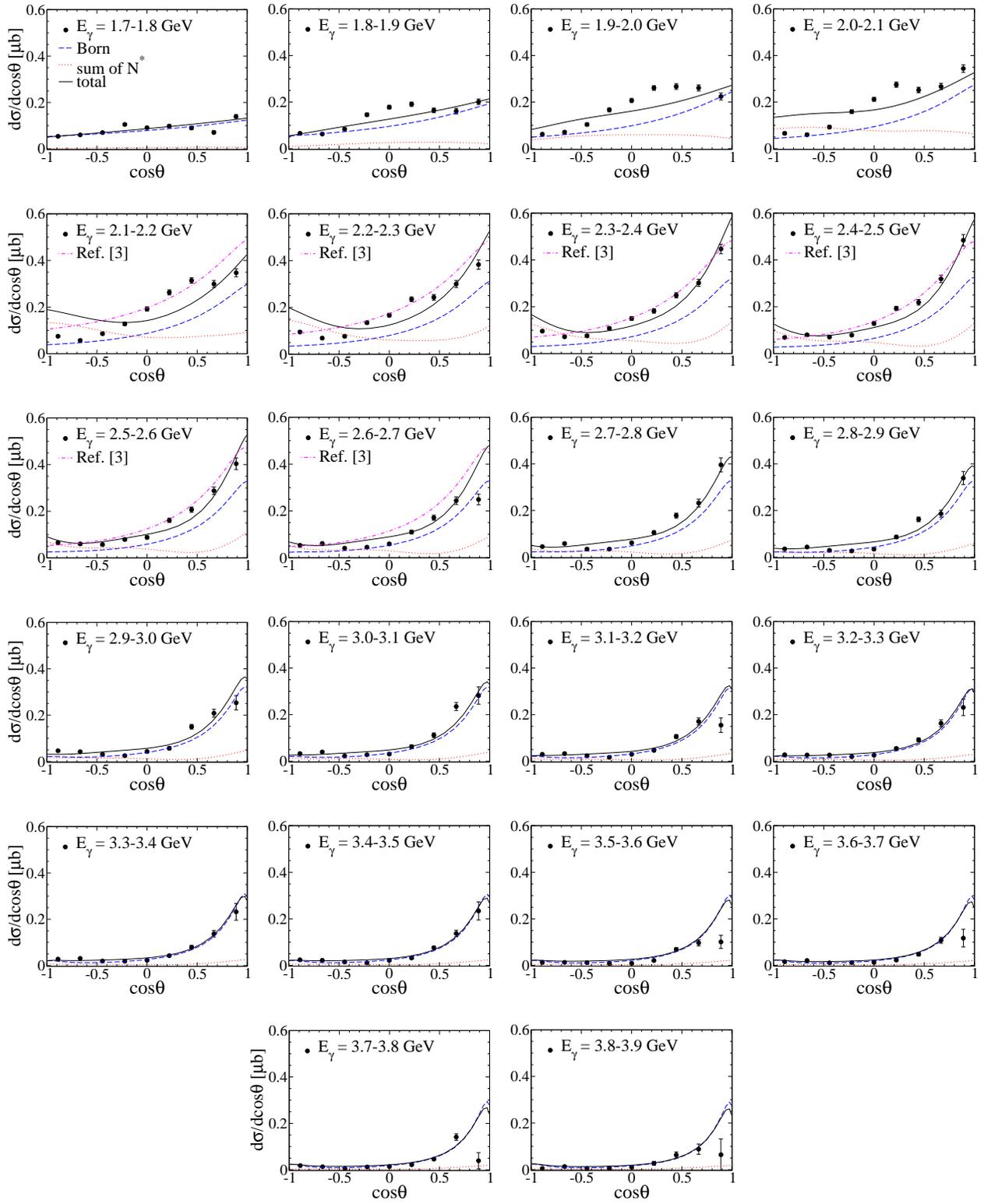

\begin{tabular}{cccc}
\hspace{-0.5em}
\includegraphics[width=4.15cm]{FIG4a.eps}\hspace{0.55em}
\includegraphics[width=3.85cm]{FIG4b.eps}\hspace{0.55em}
\includegraphics[width=3.85cm]{FIG4c.eps}\hspace{0.55em}
\includegraphics[width=3.85cm]{FIG4d.eps}\vspace{1.20em}
\\ \hspace{-0.5em} 
\includegraphics[width=4.15cm]{FIG4e.eps}\hspace{0.55em}
\includegraphics[width=3.85cm]{FIG4f.eps}\hspace{0.55em}
\includegraphics[width=3.85cm]{FIG4g.eps}\hspace{0.55em}
\includegraphics[width=3.85cm]{FIG4h.eps}\vspace{1.20em}
\\ \hspace{-0.5em}
\includegraphics[width=4.15cm]{FIG4i.eps}\hspace{0.55em}
\includegraphics[width=3.85cm]{FIG4j.eps}\hspace{0.55em}
\includegraphics[width=3.85cm]{FIG4k.eps}\hspace{0.55em}
\includegraphics[width=3.85cm]{FIG4l.eps}\vspace{1.20em}
\\ \hspace{-0.5em} 
\includegraphics[width=4.15cm]{FIG4m.eps}\hspace{0.55em}
\includegraphics[width=3.85cm]{FIG4n.eps}\hspace{0.55em}
\includegraphics[width=3.85cm]{FIG4o.eps}\hspace{0.55em}
\includegraphics[width=3.85cm]{FIG4p.eps}\vspace{1.20em}
\\ \hspace{-0.5em} 
\includegraphics[width=4.15cm]{FIG4q.eps}\hspace{0.55em}
\includegraphics[width=3.85cm]{FIG4r.eps}\hspace{0.55em}
\includegraphics[width=3.85cm]{FIG4s.eps}\hspace{0.55em}
\includegraphics[width=3.85cm]{FIG4t.eps}\vspace{1.20em}
\\ \hspace{-0.5em}
\includegraphics[width=4.15cm]{FIG4u.eps}\hspace{0.55em}
\includegraphics[width=3.85cm]{FIG4v.eps}
\end{tabular}
\caption{(Color online) Differential cross sections for the
 $\gamma p \to K^{*+}\Lambda$ reaction as a function of $\cos\theta$
in the range of $1.7\,\mathrm{GeV} \le E_\gamma \le 3.9\, \mathrm{GeV}$.
The notations are the same as in the left panel of Fig.~\ref{FIG3}. }
\label{FIG4}
\end{figure}

In Fig.~\ref{FIG4}, the differential cross sections are plotted as a 
function of $\cos\theta$  in the range of $1.7\,\mathrm{GeV} \le
E_\gamma \le 3.9\, \mathrm{GeV}$. The effects of the $N^*$
resonances seem to be negligible in the vicinity of the threshold
energy as shown in the first panel of Fig.~\ref{FIG4}. However, as the
photon energy increases, the $N^*$ resonances come into play. Apart
from the structures of a broad bump in the experimental data in the
range $1.8\,\mathrm{GeV}\le E_\gamma \le 2.3\,\mathrm{GeV}$, the
present results are in good agreement with the data in general. 
Experimentally, the differential cross sections in the forward
direction starts to increase as $E_\gamma$ does. This feature is
qualitatively explained by the Born terms but can be described
quantitatively only by including the $N^*$ resonances.  
The experimental data in the forward direction remain almost constant
with little energy dependence in the range of
$3.2\,\mathrm{GeV} \le E_\gamma \le 3.5\,\mathrm{GeV}$
then start to fall off drastically above 3.5 GeV. 
The present model is not able to describe this behavior of the data. 
Considering the fact that the present approach of effective Lagrangians 
is built for lower $E_\gamma$ regions, one has to take into account more 
degrees of freedom or a more sophisticated theoretical method to
explain the $\gamma p\to K^{*+}\Lambda$ at higher photon energies. 
On the other hand, as shown in some energy range, i.e. 
$2.1\,\mathrm{GeV} \le E_\gamma \le 2.6\, \mathrm{GeV}$,  
we find with the main contribution of $D_{13}(2080)$~\cite{Kim:2011rm}
that the theoretical calculations slightly overestimate the CLAS data,  
which is expected from Fig.~\ref{FIG2}. We want to mention that the
role of the $N^*$ resonances for the $\gamma N \rightarrow K^*\Lambda$
reaction look very different from that for $\gamma p \rightarrow
K^*\Sigma$, where the $N^*$ resonances are almost negligible. We refer
to Refs.~\cite{SHKim2:2013,SHKim3:2013} for details. 

\begin{figure}[ht]
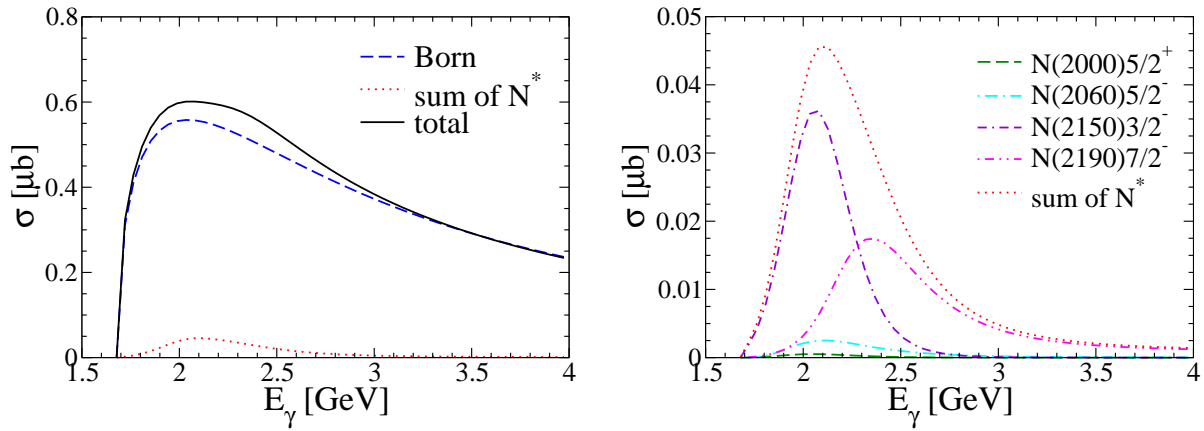

\vspace{1.5em}
\begin{tabular}{cc}
\includegraphics[scale=0.3]{FIG5a.eps} \;\;\;\;
\includegraphics[scale=0.3]{FIG5b.eps}
\end{tabular}
\caption{(Color online) The results of the total
  cross sections for the  $\gamma n \to K^{*0}\Lambda$ reaction in the
  left panel. The solid curve draws the total contribution of all
  diagrams, whereas the dashed one shows that of the Born terms except
  for the $N^*$ resonances. The dotted curve depicts the
  contribution of the $N^*$ resonances to the total cross
  section. The right panel illustrates each contribution
  of the $N^*$ resonances.}
\label{FIG5}
\end{figure}
\begin{figure}[ht]
\vspace{0.5em}
\begin{tabular}{cccc}
\hspace{-0.5em}
\includegraphics[width=4.15cm]{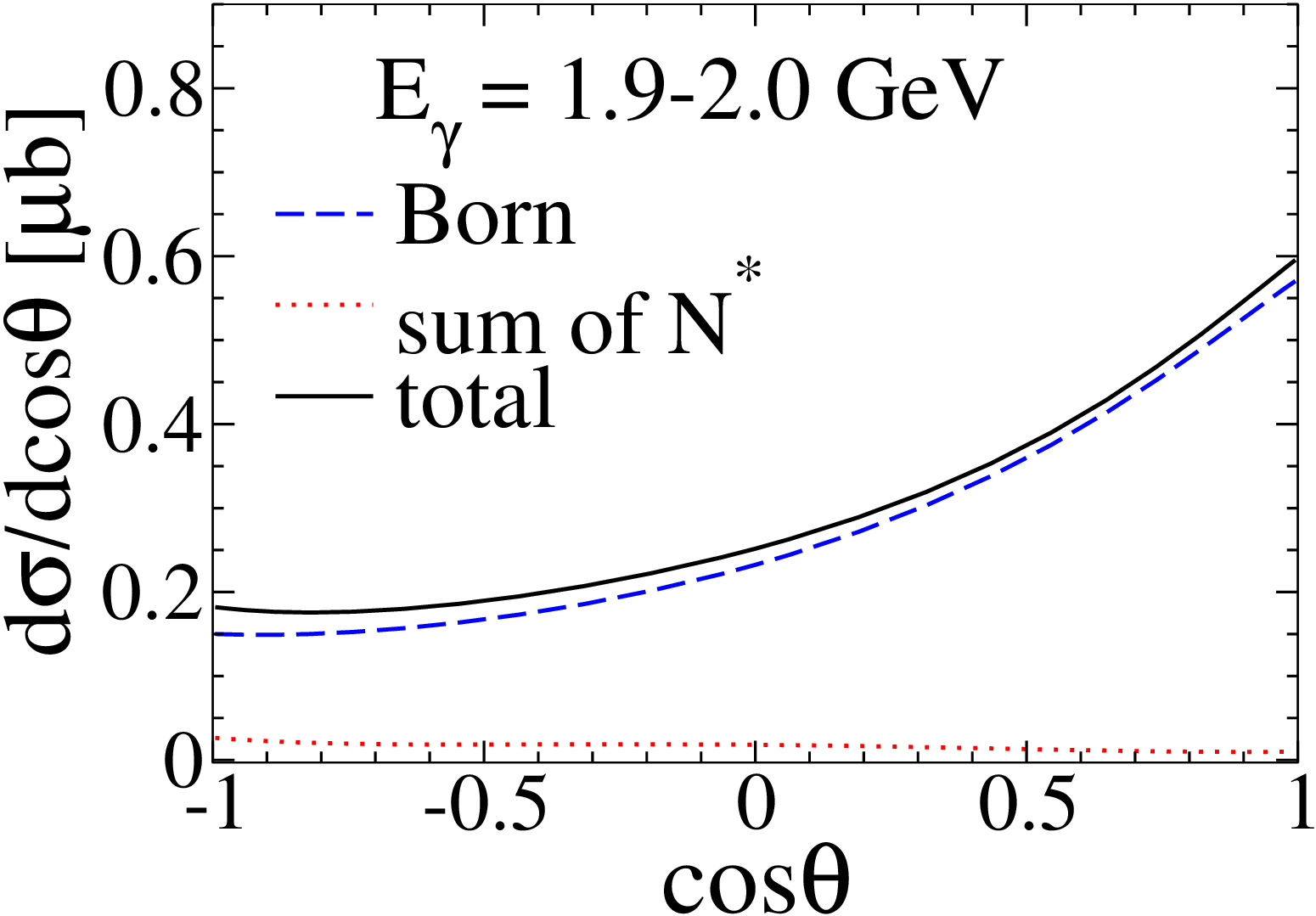}\hspace{0.55em}
\includegraphics[width=3.85cm]{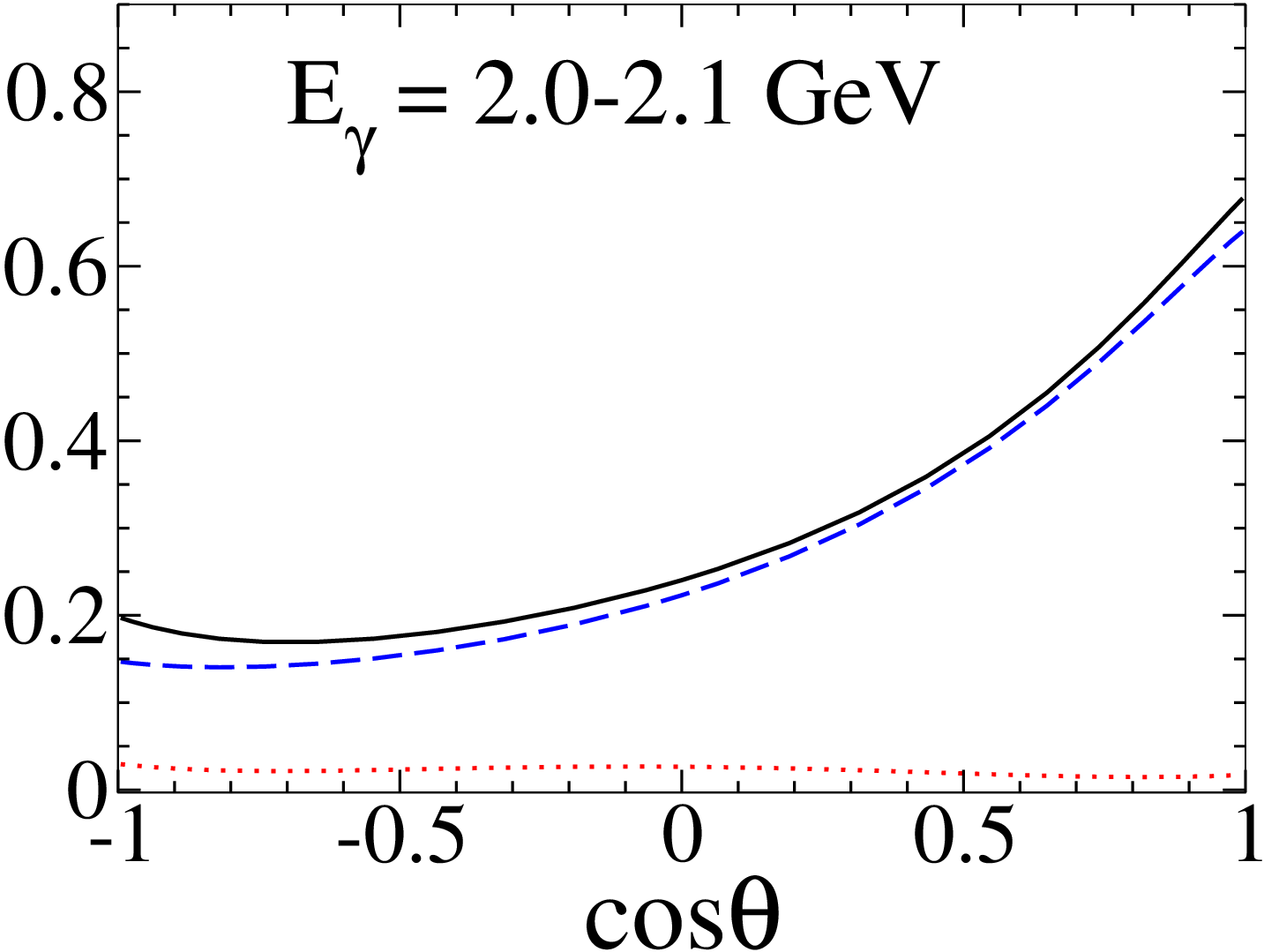}\hspace{0.55em}
\includegraphics[width=3.85cm]{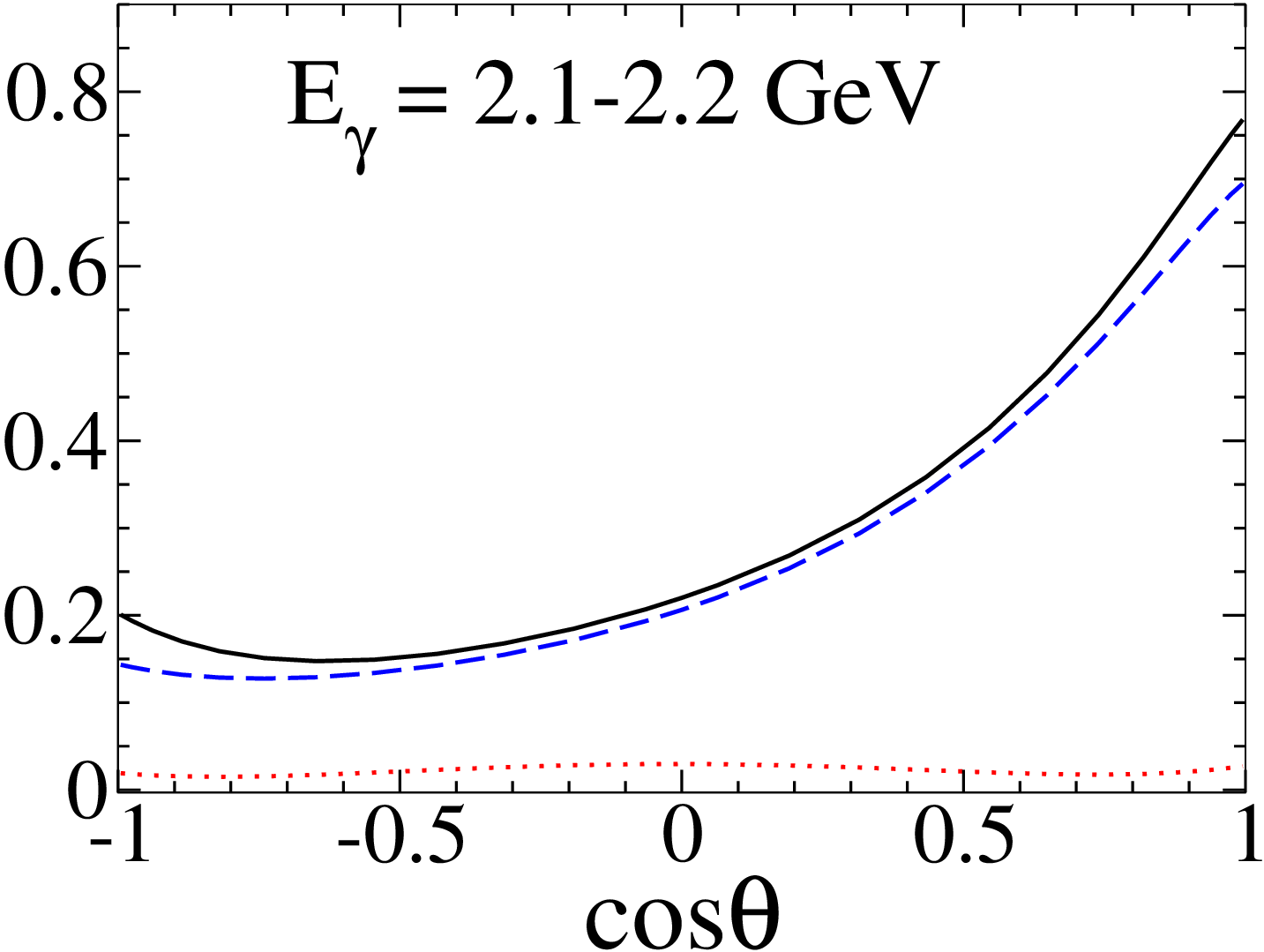}\hspace{0.55em}
\includegraphics[width=3.85cm]{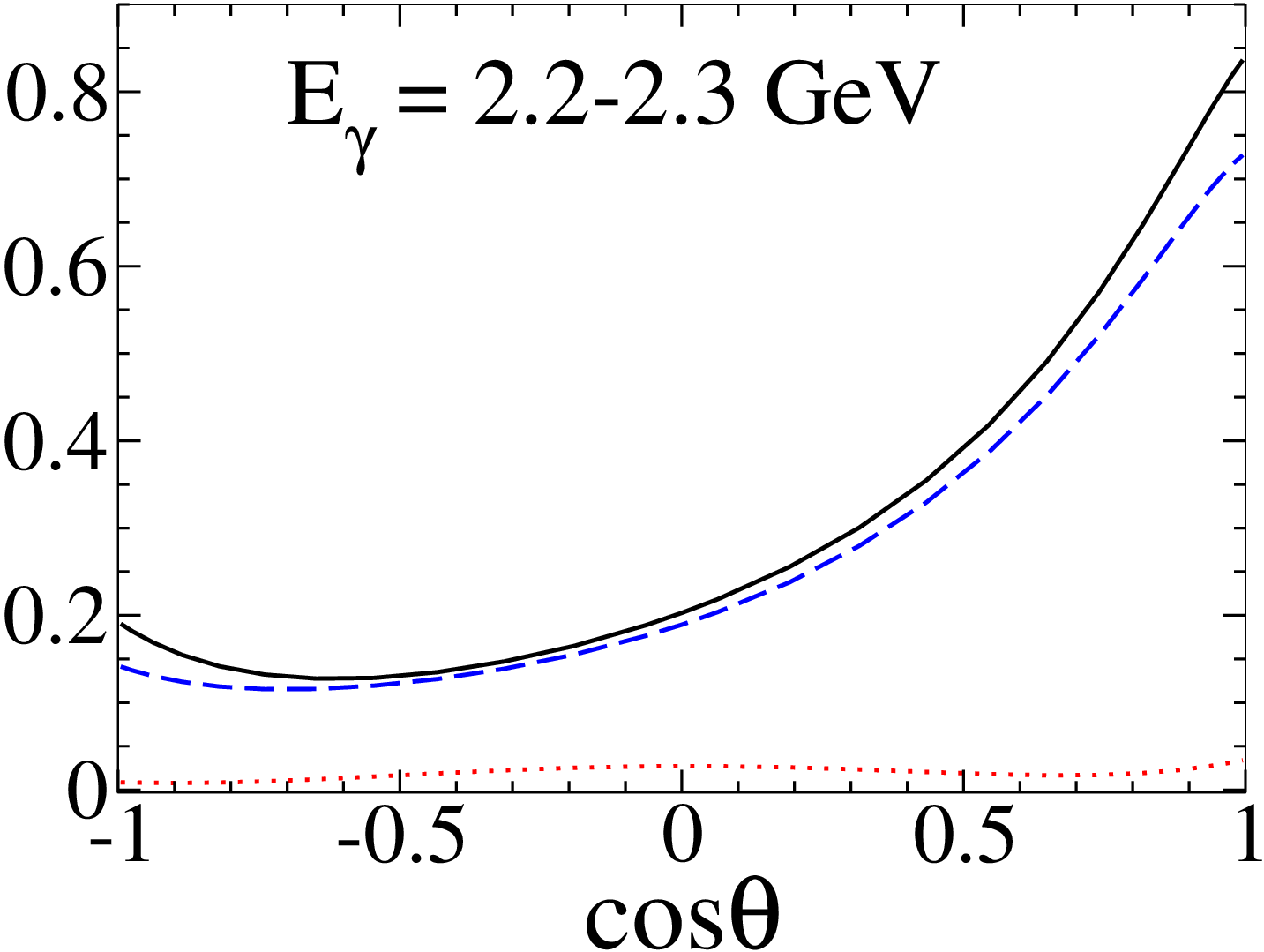}\vspace{1.20em}
\\ \hspace{-0.5em}
\includegraphics[width=4.15cm]{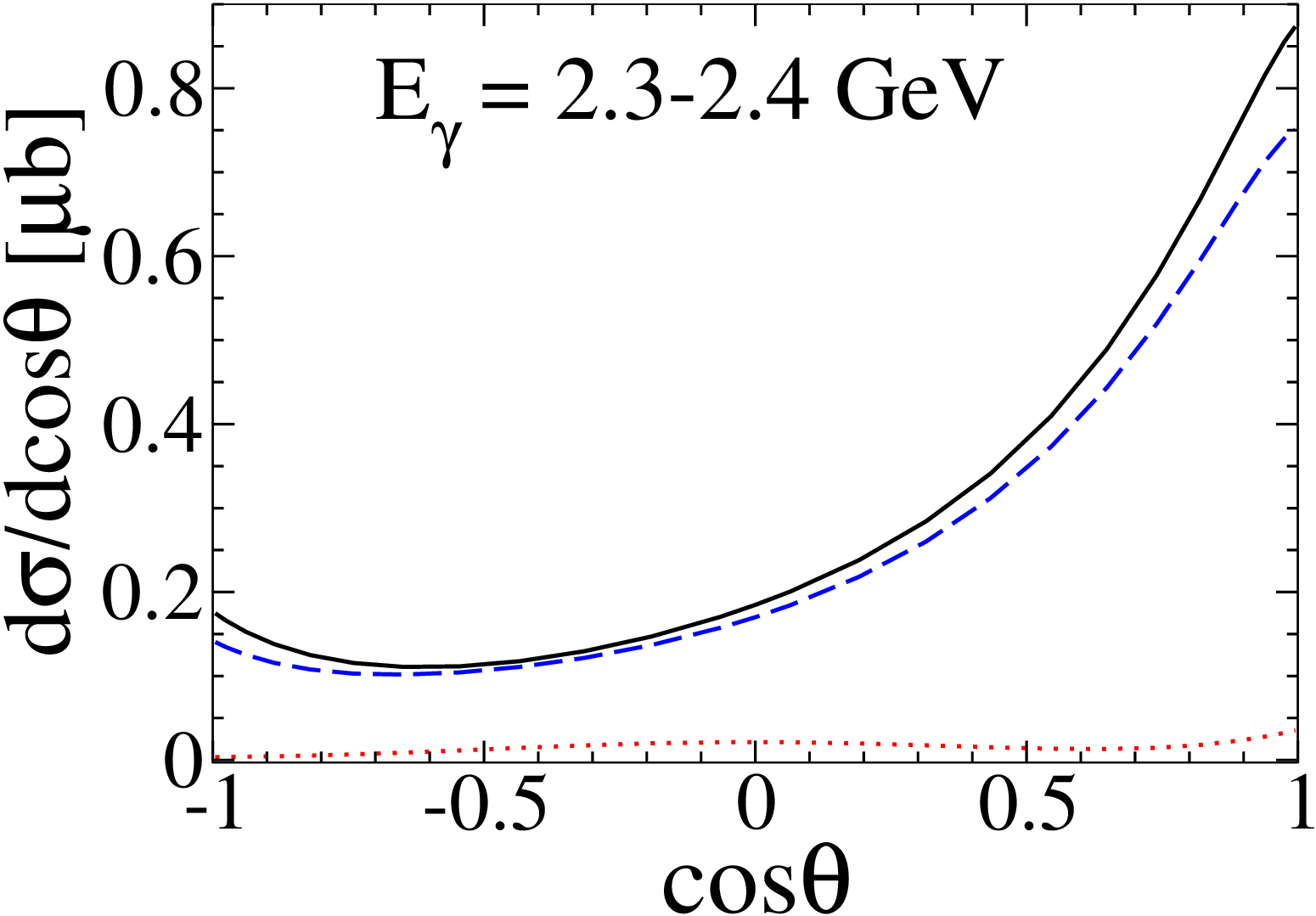}\hspace{0.55em}
\includegraphics[width=3.85cm]{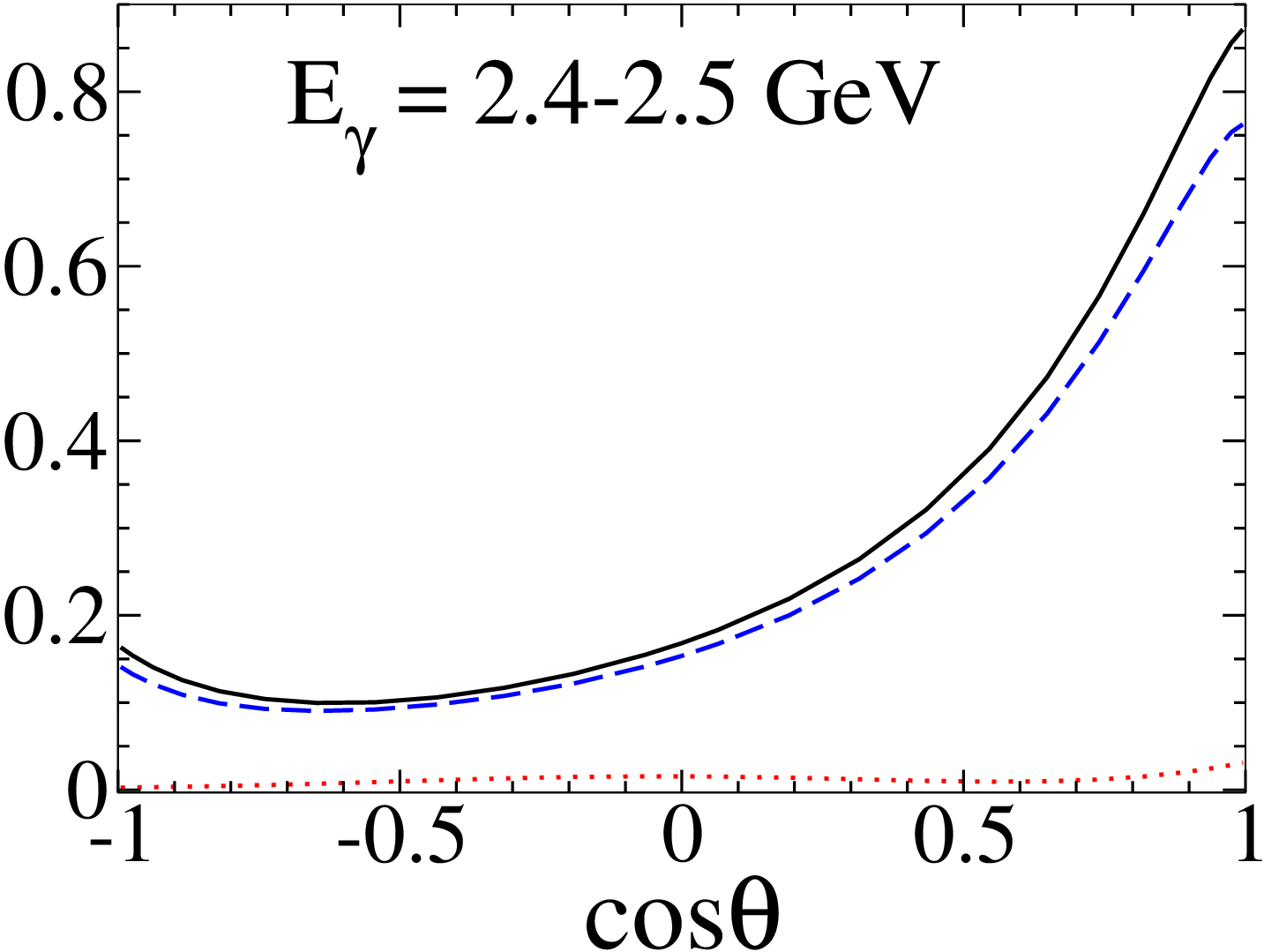}\hspace{0.55em}
\includegraphics[width=3.85cm]{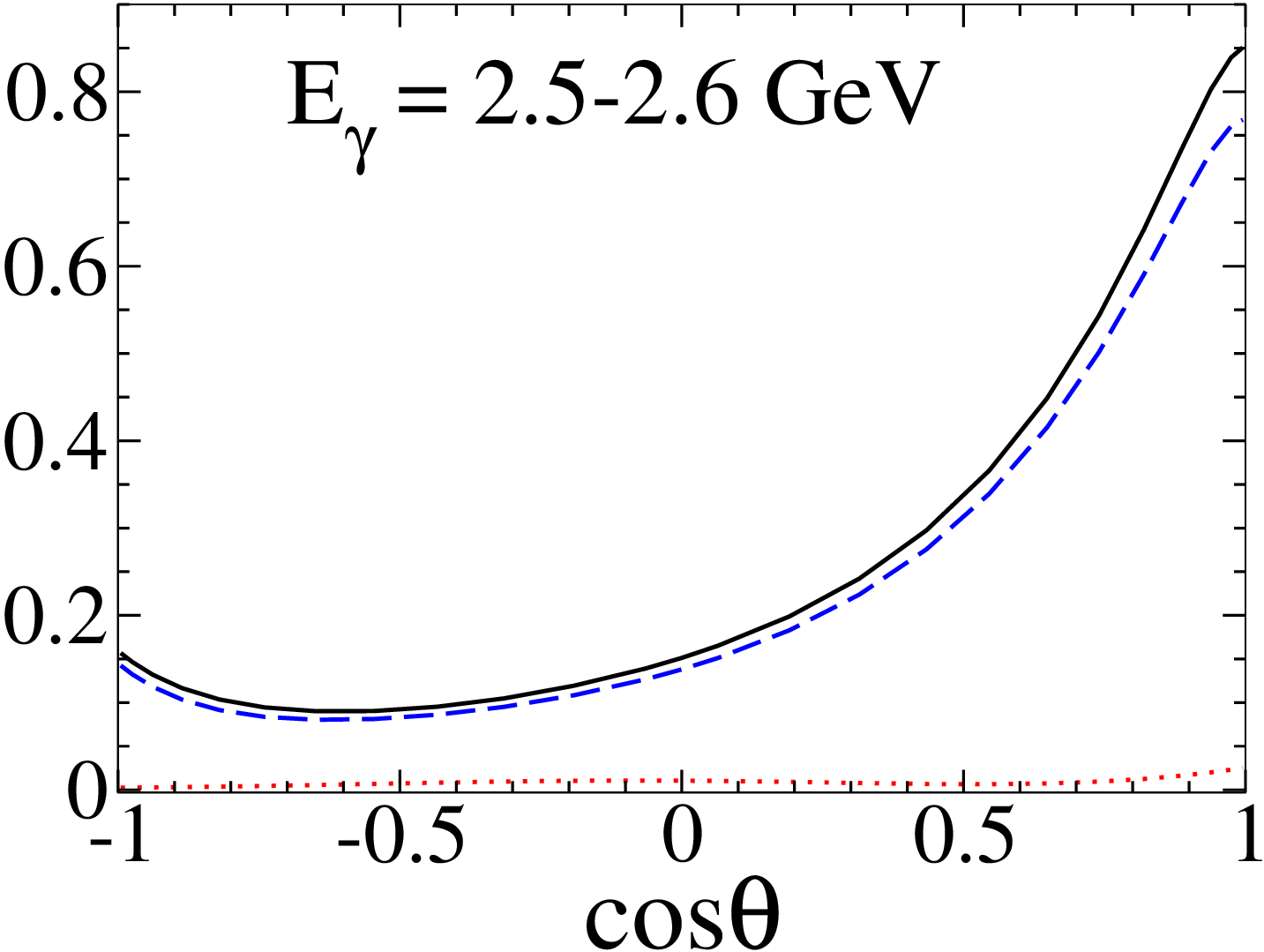}\hspace{0.55em}
\includegraphics[width=3.85cm]{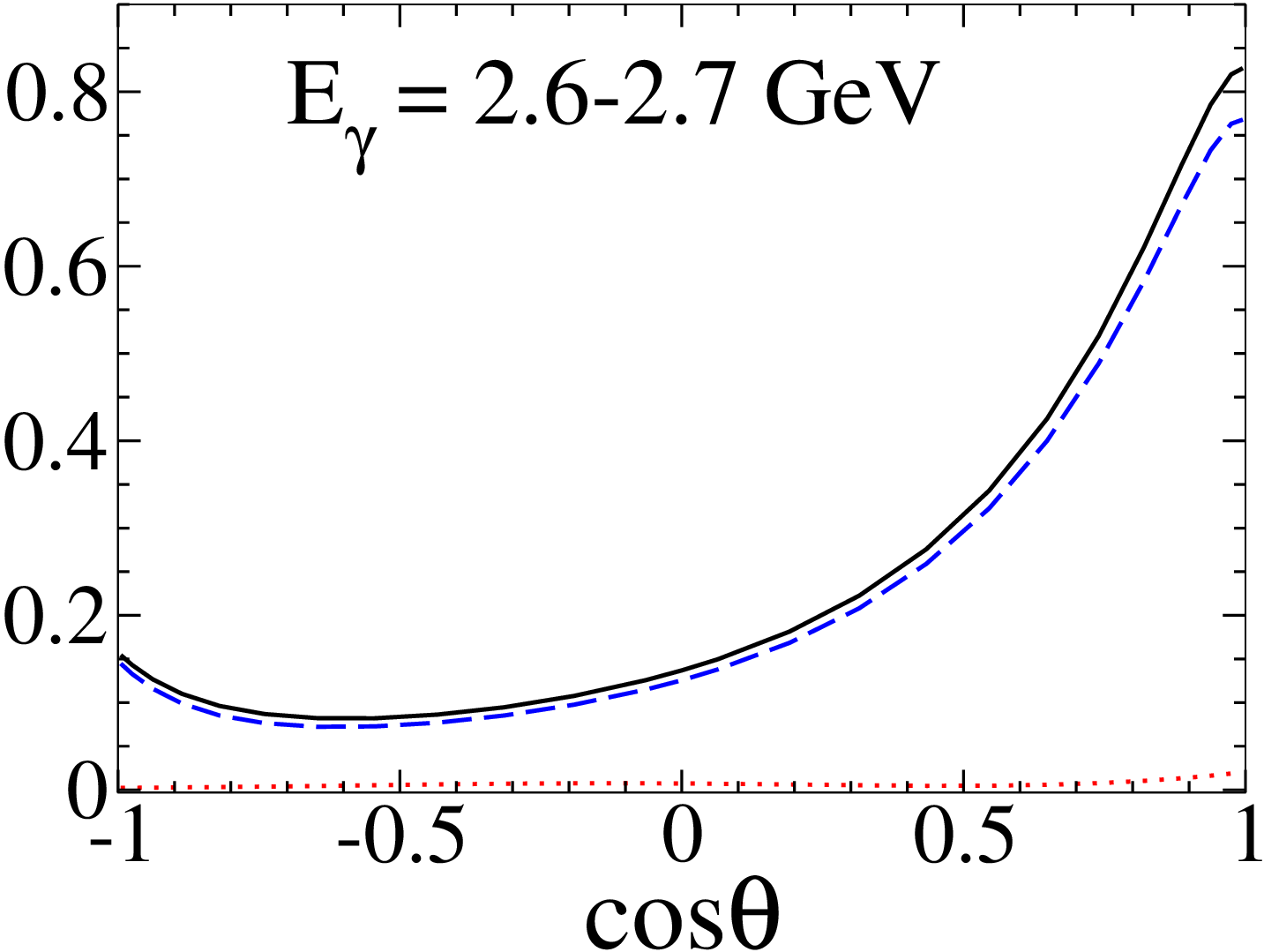}
\end{tabular}
\caption{(Color online) Differential cross sections for the
 $\gamma n \to K^{*0}\Lambda$ reaction as a function of $\cos\theta$
in the range of $1.9\,\mathrm{GeV} \le E_\gamma \le 2.7\, \mathrm{GeV}$.
The notations are the same as in the left panel of Fig.~\ref{FIG5}.}
\label{FIG6}
\end{figure}
In the left panel of Fig.~\ref{FIG5}, we predict the total cross
section for the $\gamma n \to K^{*0}\Lambda$ reaction. The neutral
charge of the $K^{*0}$ makes the $K^*$ exchange and the contact term
absent in this reaction. Nevertheless, the magnitude of this total
cross section is quite larger than that of the charged process $\gamma
p \to K^{*+}\Lambda$ because of the large neutral coupling constant of 
the $\gamma KK^*$ interaction, as discussed in detail in
Ref.~\cite{Kim:2011rm}. Thus, the main contribution to the total cross
section for the $\gamma n \to K^{*0}\Lambda$ reaction arises from the
$K$ exchange. Moreover, the effects of the $N^*$ resonances are
almost marginal for the neutral process. Each contribution of the four
$N^*$ resonances drawn in the right panel of Fig.~\ref{FIG5}.
Figure~\ref{FIG6} depicts the differential cross section as a function
of $\cos\theta$ with $E_\gamma$ varied from 1.9 GeV to 2.7 GeV. 
The experimental data for the $\gamma n \to K^{*0}\Lambda$ reaction
will soon appear.

We now want to discuss the polarization
observables~\cite{Fasano:1992es, Pichowsky:1994gh, Titov:1998bw},
which provide crucial information on the helicity amplitudes and spin 
structure of a process. To define the polarization observables, the 
reaction takes place in the $x-z$ plane with the photon beam .
We first start with the single polarization
observables. Since we consider also the double polarization
observables, we will follow the notation for the polarized
differential cross sections defined in Ref.~\cite{Titov:1998bw} 
\begin{equation}
d\sigma(B,T;R,V) \;=\;
\frac{d\sigma}{d\Omega}(B,T;R,V),   
\label{eq:11} 
\end{equation}
where $B$, $T$, $R$, $V$ denote the polarizations
of the photon beam ($B$), the target nucleon ($T$), the recoil
$\Lambda$ ($R$), and the produced $K^*$ vector meson ($V$),
respectively, involved in $\gamma N\to K^* \Lambda$ process.
According to the notation defined in Eq.~(\ref{eq:11}), 
we define the photon-beam asymmetry ($\Sigma_x$), the target 
asymmetry ($T_y$), and the recoil asymmetry ($P_y$) as 
\begin{eqnarray}
\Sigma_x &=&
{\frac{d\sigma{(\perp,U;U,U)}-d\sigma{(\parallel,U;U,U)}}
{d\sigma{(\perp,U;U,U)}+d\sigma{(\parallel,U;U,U)}}} ,       \cr
T_y &=&
{\frac{d\sigma{(U,y;U,U)}-d\sigma{(U,-y;U,U)}}
{d\sigma{(U,y;U,U)}+d\sigma{(U,-y;U,U)}}} ,                  \cr 
P_y &=& 
{\frac{d\sigma{(U,U;y,U)}-d\sigma{(U,U;-y,U)}}
{d\sigma{(U,U;y,U)}+d\sigma{(U,U;-y,U)}}}  ,                  
\end{eqnarray}
where $\parallel$ and $\perp$ denote the linear polarizations of the
photon along the direction of the $\bm x$ and $\bm y$ axes, 
respectively.
$y$ and $-y$ represent the polarization states of the $N$ ($\Lambda$),
which lie in the direction of the $\bm y$ and $-\bm y$ axes , respectively.
The $U$ means that the corresponding particle state is
unpolarized. These three asymmetries satisfy the following collinear
condition 
\begin{equation}
  \label{eq:13}
\Sigma_x = T_y = P_y = 0 \mbox{ at } \cos\theta = \pm 1.  
\end{equation}

\begin{figure}[ht]
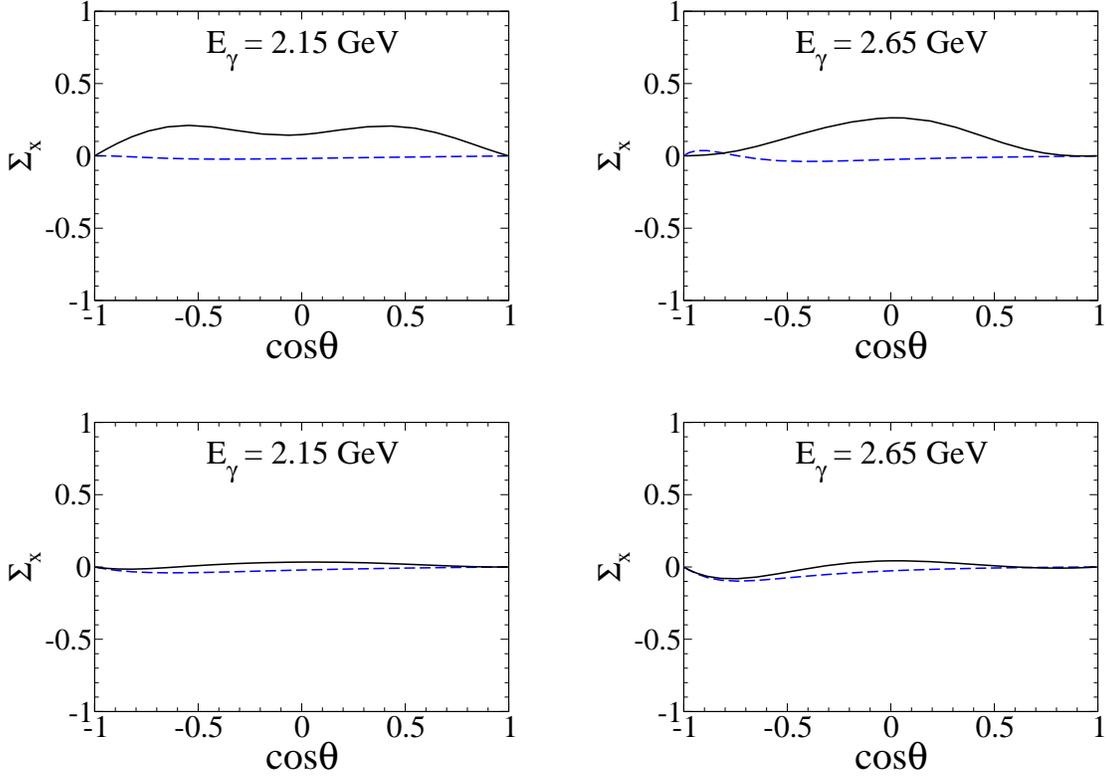

\begin{tabular}{cc}
\includegraphics[width=6.75cm]{FIG7a.eps}\hspace{3.0em}
\includegraphics[width=6.75cm]{FIG7b.eps}\vspace{1.8em}\\
\includegraphics[width=6.75cm]{FIG7c.eps}\hspace{3.0em}
\includegraphics[width=6.75cm]{FIG7d.eps}
\end{tabular}
\caption{(Color online) The photon-beam asymmetries as functions of 
  $\cos\theta$ with two different photon energies, $E_\gamma=2.15
  \,\mathrm{GeV}$ and $E_\gamma=2.65  \,\mathrm{GeV}$.
In the upper panel, we draw the $\Sigma_x$ for the $\gamma p\to
K^{*+}\Lambda$ reaction, while in the lower panel we do for the
$\gamma n\to K^{*0}\Lambda$ reaction. The solid curves represent the
total results including the $N^*$ resonances, whereas the dashed ones
show those without them.}    
\label{FIG7}
\end{figure}
\begin{figure}[ht]
\begin{tabular}{cc}
\includegraphics[width=6.75cm]{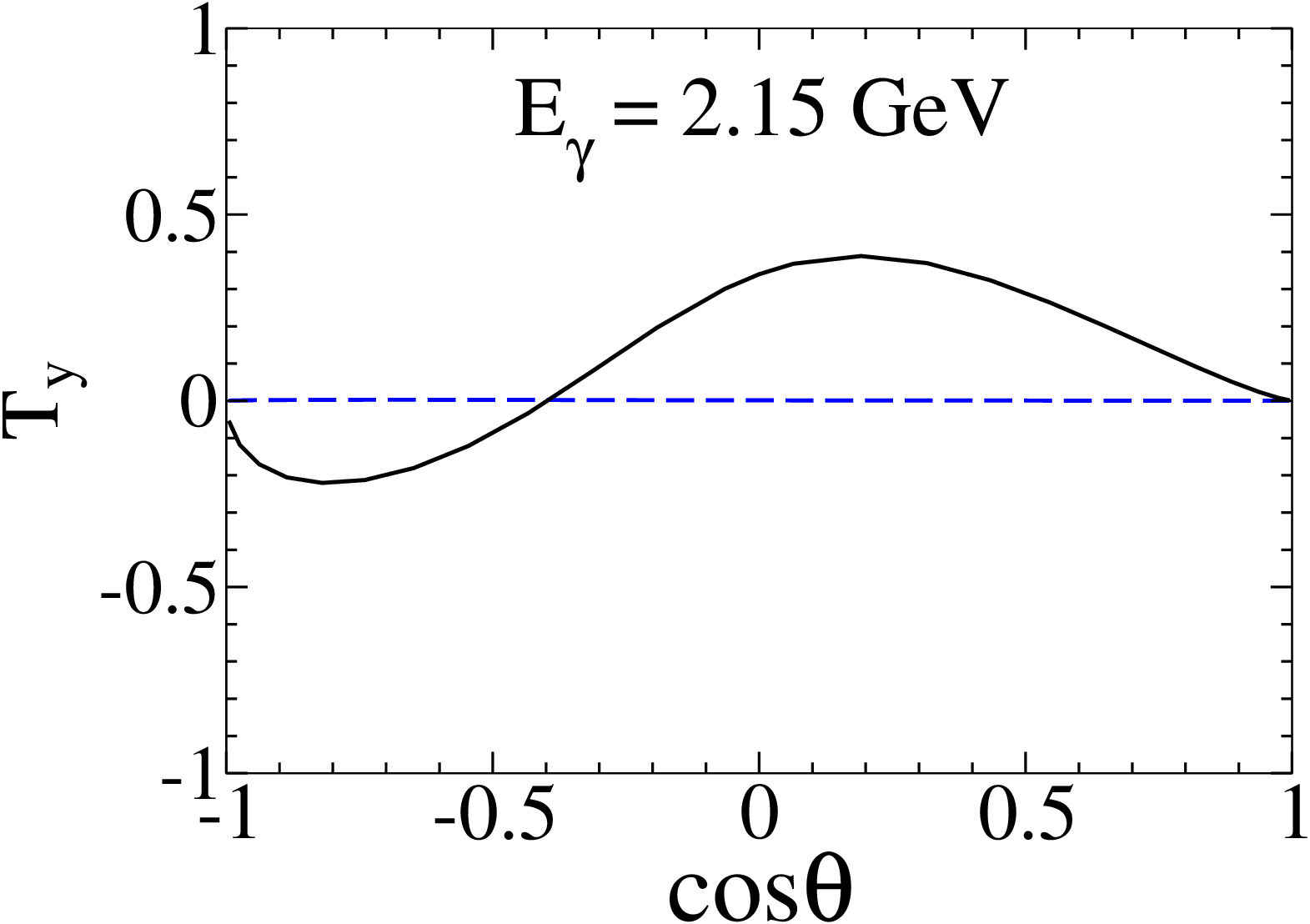}\hspace{3.0em}
\includegraphics[width=6.75cm]{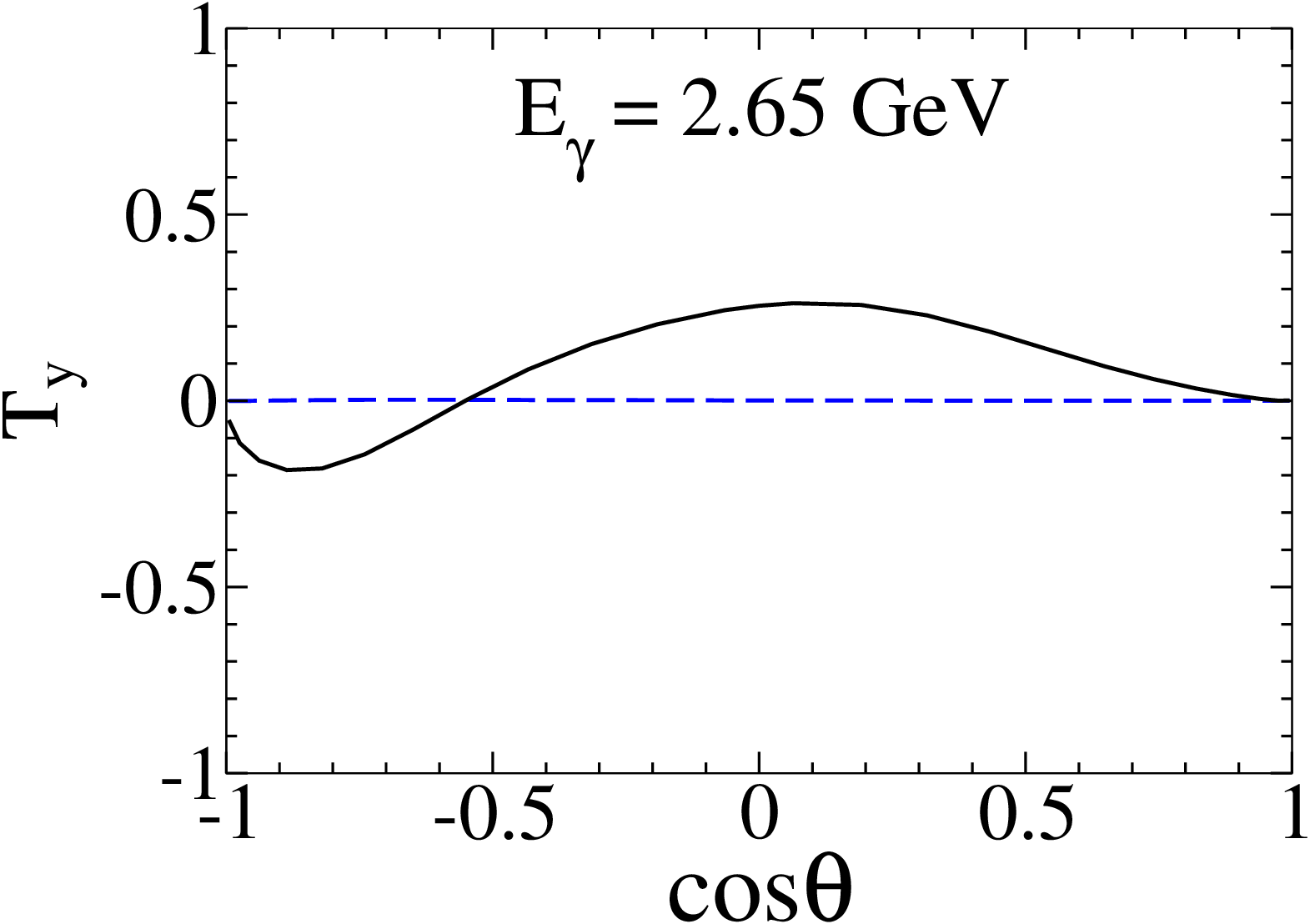}\vspace{1.7em}\\
\includegraphics[width=6.75cm]{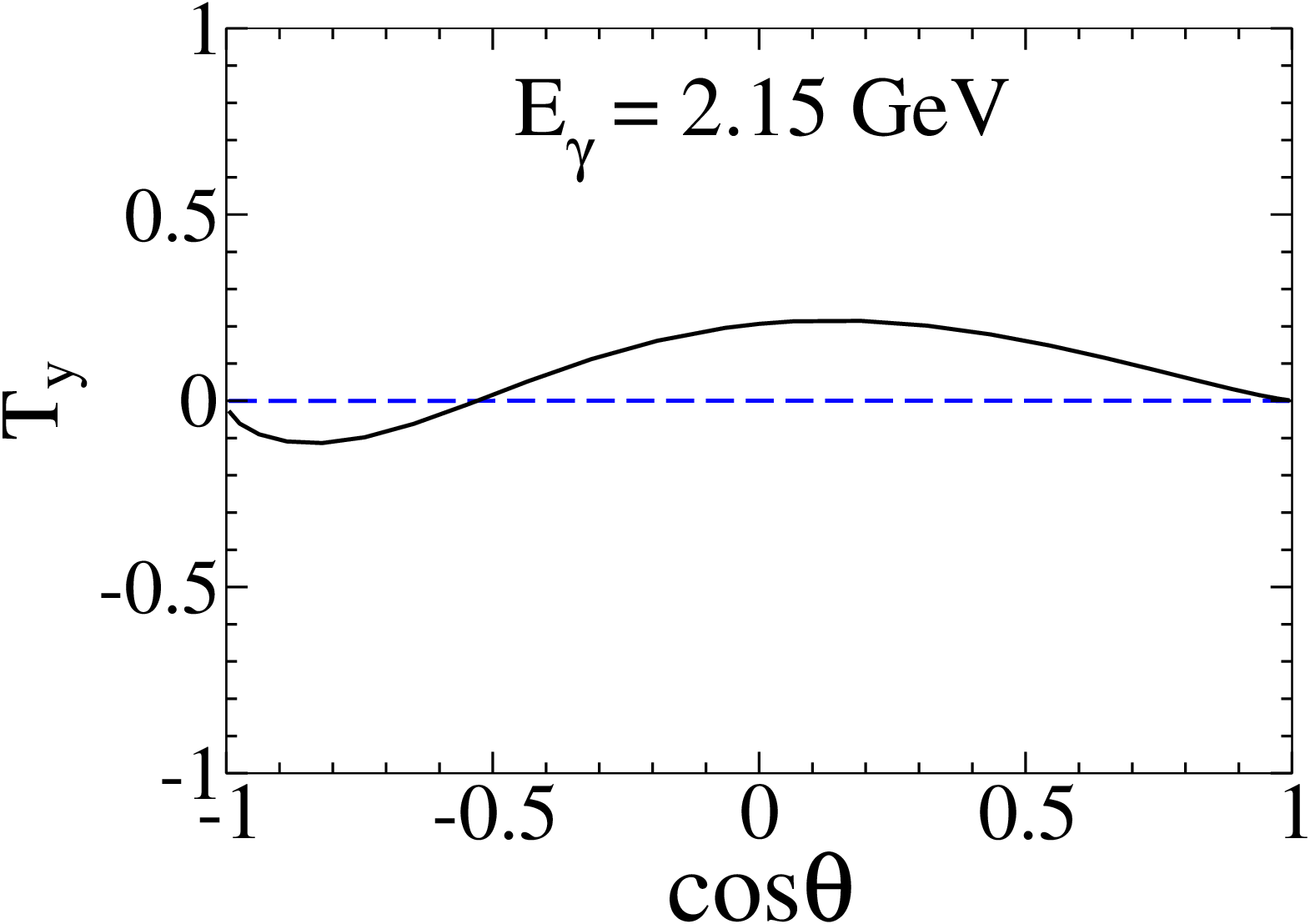}\hspace{3.0em}
\includegraphics[width=6.75cm]{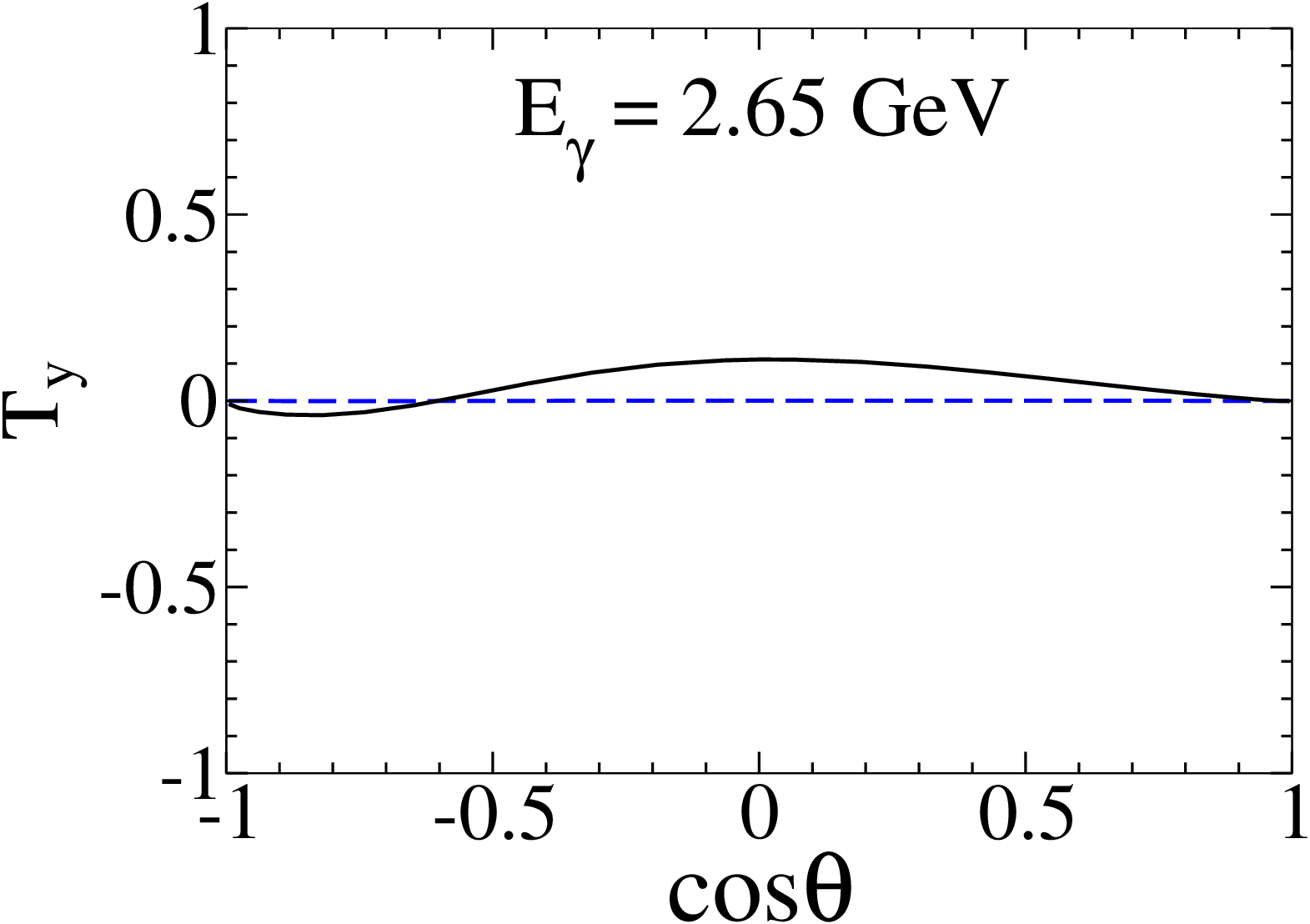}
\end{tabular}
\caption{(Color online) The target asymmetries as functions of 
  $\cos\theta$ with two different photon energies, $E_\gamma=2.15
  \,\mathrm{GeV}$ and $E_\gamma=2.65  \,\mathrm{GeV}$.
In the upper panel, we draw the $T_y$ for the $\gamma p\to
K^{*+}\Lambda$ reaction, while in the lower panel we do for the
$\gamma n\to K^{*0}\Lambda$ reaction.
Notations are the same as in Fig.~\ref{FIG7}.}
\label{FIG8}
\end{figure}
\begin{figure}[ht]
\begin{tabular}{cc}
\includegraphics[width=6.75cm]{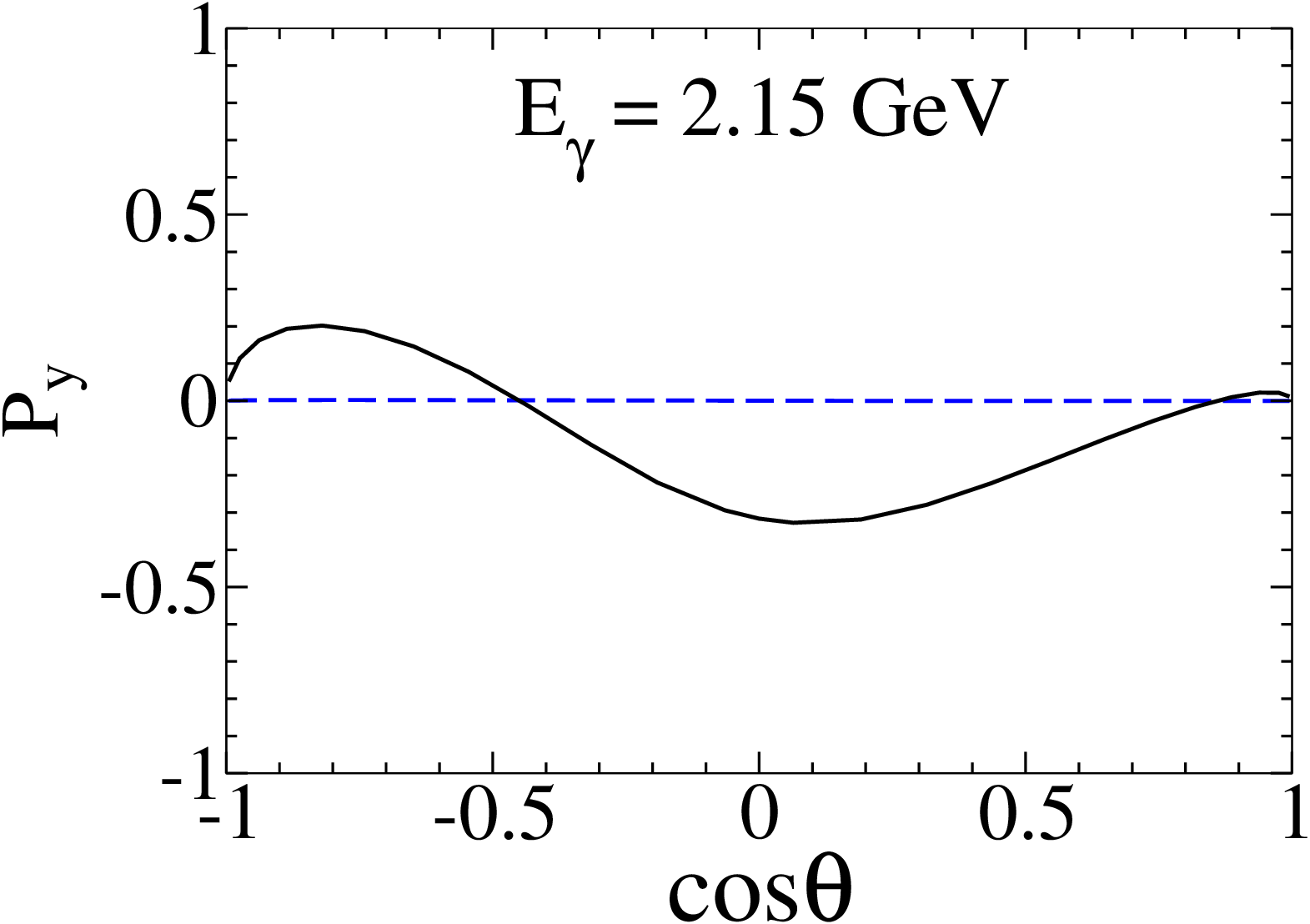}\hspace{3.0em}
\includegraphics[width=6.75cm]{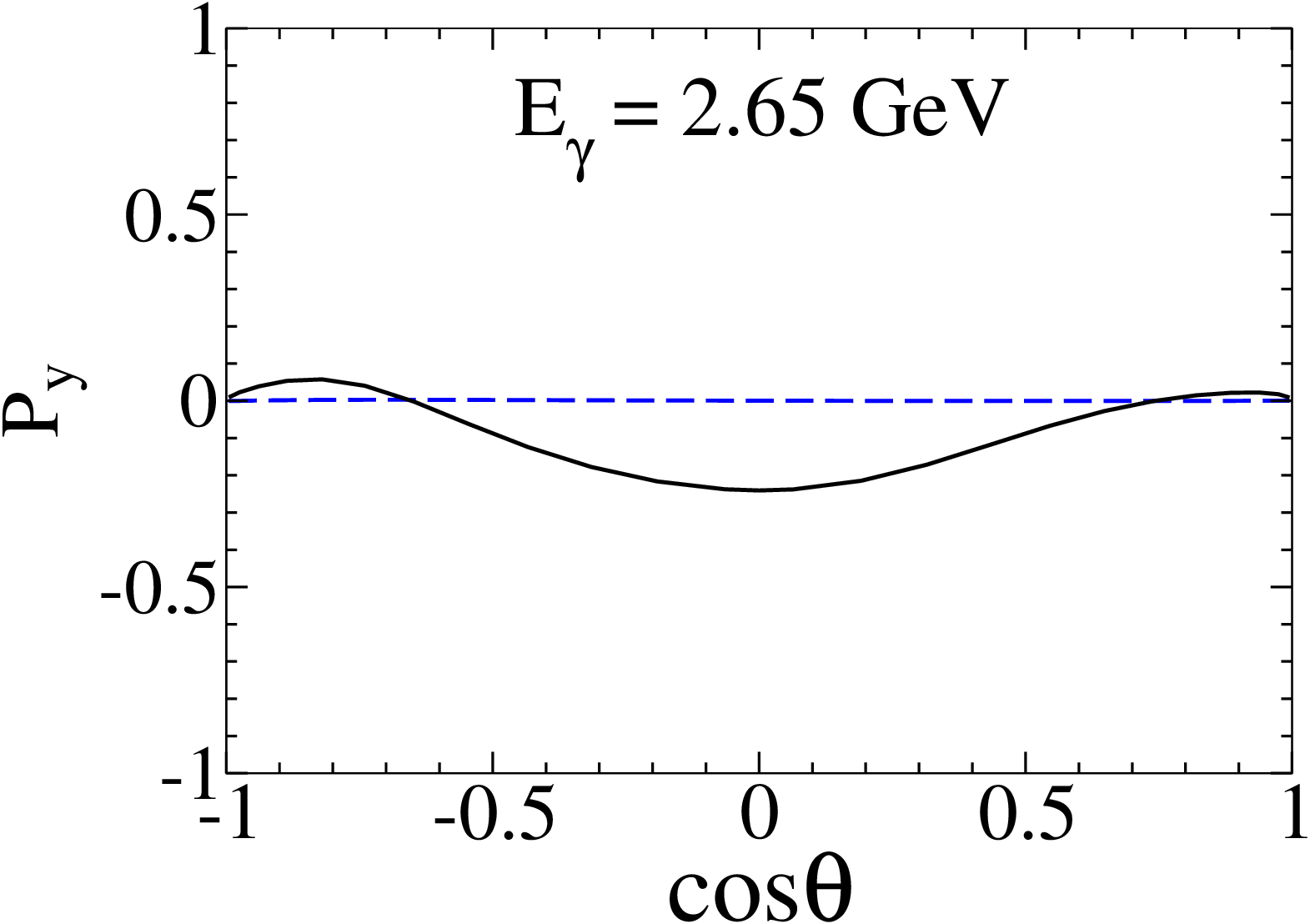}\vspace{1.7em}\\
\includegraphics[width=6.75cm]{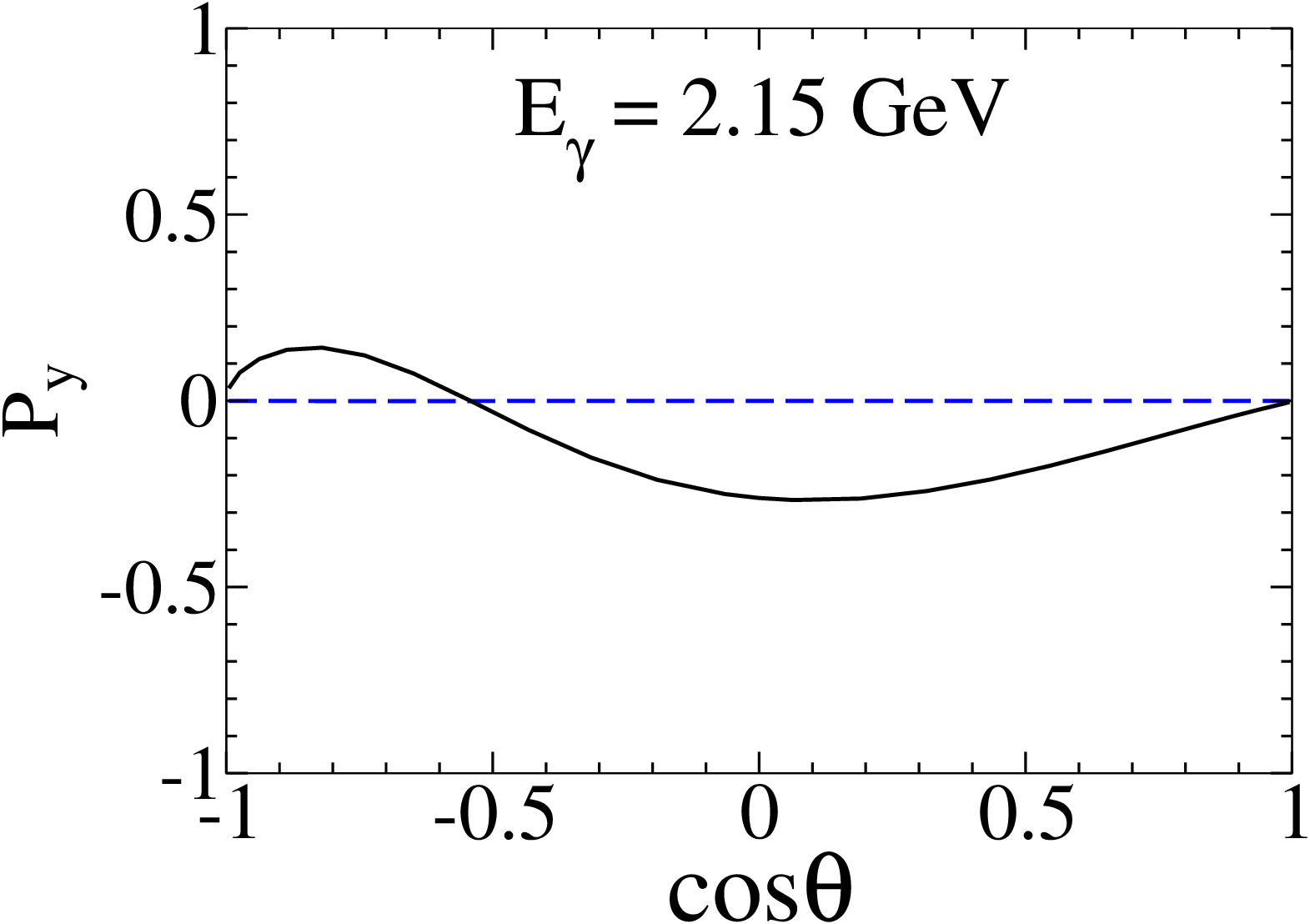}\hspace{3.0em}
\includegraphics[width=6.75cm]{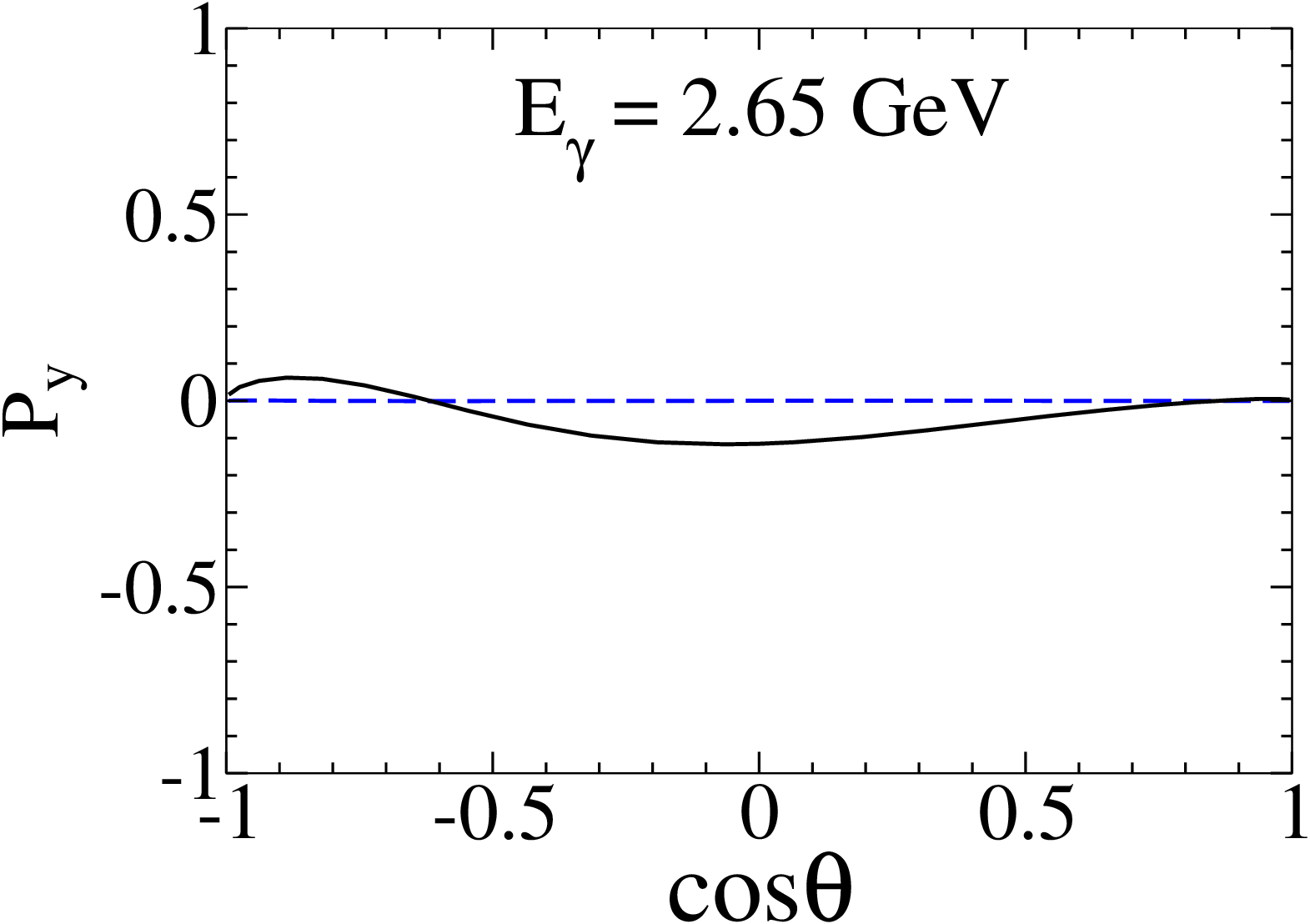}
\end{tabular}
\caption{(Color online) The recoil asymmetries as functions of 
  $\cos\theta$ with two different photon energies, $E_\gamma=2.15
  \,\mathrm{GeV}$ and $E_\gamma=2.65  \,\mathrm{GeV}$.
In the upper panel, we draw the $P_y$ for the $\gamma p\to
K^{*+}\Lambda$ reaction, while in the lower panel we do for the
$\gamma n\to K^{*0}\Lambda$ reaction. 
Notations are the same as in Fig.~\ref{FIG7}.
}
\label{FIG9}
\end{figure}
The upper panel of Fig.~\ref{FIG7} depicts the results of the
photon-beam asymmetries for the charged process $\gamma p\to
K^{*+}\Lambda$ at two different photon energies,
$E_\gamma=2.15\,\mathrm{GeV}$ and $E_\gamma = 2.65\,\mathrm{GeV}$. As
already discussed in Ref.~\cite{Kim:2011rm}, the beam asymmetry is
almost compatible with zero without the $N^*$ resonances. Including
them, we find that $\Sigma_x$ becomes positive and has broad bump
structures. Thus, the measurement of the beam symmetry can already
tell whether the $N^*$ resonances are indeed important in
understanding the production mechanism of the $\gamma p\to
K^{*+}\Lambda$ reaction. In the lower panel of Fig.~\ref{FIG7}, we
draw the results of the $\Sigma_x$ for the neutral $\gamma n\to
K^{*0}\Lambda$ reaction. It is interesting to see that the effects of
the $N^*$ resonances turn out to be rather small in this case. We
already saw that their contribution to the total and differential
cross sections are marginal, since the contribution of the $K$ meson
exchange governs the $\gamma n\to K^{*0}\Lambda$ reaction. By the
same token, the effects of the $N^*$ resonances seem to be suppressed
in the beam asymmetries for the neutral process. 

In Figs.~\ref{FIG8} and \ref{FIG9}, 
the results of the target and recoil asymmetries are
drawn, respectively. As in Fig.~\ref{FIG7}, the upper panel is for the
$\gamma p\to K^{*+}\Lambda$ reaction and the lower panel corresponds
to $\gamma n\to K^{*0}\Lambda$ reaction, respectively. The dependence of
the $T_y$ on $\cos\theta$ is distinguished from that of the
beam asymmetry. The values of the $T_y$ become positive from the very
forward angle till the backward angle, and then turn negative around
$\cos\theta=-0.5\, (\theta=120^\circ)$. In the case of 
the recoil asymmetries, the results are just opposite to those of
the target asymmetries. Both the target and recoil asymmetries become
smaller in magnitude, as the photon energy increases, while the form
of the dependence on the scattering angle is kept. As found in the
results of the differential cross sections, the $N^*$ resonances
mainly explain the production mechanism in the vicnity of the
threshold energy. This can be seen also in the single spin
polarization obervables. 

We now discuss the double polarization asymmetries. In fact, there are
many different polarization observables in the vector meson 
photoproduction. Here, we will consider only some of the double
polarization asymmetries, which are defined as follows:
\begin{eqnarray}
 C^{\mathrm{BT}}_{zz} &=&
\frac{d\sigma(r,z;U,U)-d\sigma(r,-z;U,U)}
{d\sigma(r,z;U,U) + d\sigma(r,-z;U,U)} ,         \cr
C^{\mathrm{BR}}_{zz} &=&
\frac{d\sigma(r,U;z,U)-d\sigma(r,U;-z,U)}
{d\sigma(r,U;z,U) + d\sigma(r,U;-z,U)}  ,        \cr
C^{\mathrm{TR}}_{zz} &=&
\frac{d\sigma(U,z;z,U)-d\sigma(U,z;-z,U)}
{d\sigma(U,z;z,U) + d\sigma(U,z;-z,U)}  ,        \cr
C^{\mathrm{TV}}_{zz} &=&
\frac{d\sigma(U,z;U,r)-d\sigma(U,-z;U,r)}
{d\sigma(U,z;U,r) + d\sigma(U,-z;U,r)}  ,       \cr
C^{\mathrm{RV}}_{zz} &=&
\frac{d\sigma(U,U;z,r)-d\sigma(U,U;-z,r)}
{d\sigma(U,U;z,r) + d\sigma(U,U;-z,r)},         
\end{eqnarray}
where $r$ denotes the circularly polarized photon beam 
(produced vector meson) with helicity $+1$.
$\pm z$ stands for the direction of
the N ($\Lambda$) polarization. The $C_{zz}^{BT}$,
$C_{zz}^{BR}$, $C_{zz}^{TR}$, $C_{zz}^{TV}$, and
$C_{zz}^{RV}$ are respectively called the beam-target (BT)
asymmetry, the beam-recoil (BR) asymmetry, the target-recoil (TR)
asymmetry, the target-vector-meson (TV) asymmetry, and the 
recoil-vector-meson (RV) asymmetry. We will now see that the effects of 
the $N^*$ resonances are even more dramatic, in particular, in the case 
of the $\gamma p\to K^{*+}\Lambda$ reaction.   

\begin{figure}[ht]
\begin{tabular}{cc}
\includegraphics[width=6.5cm]{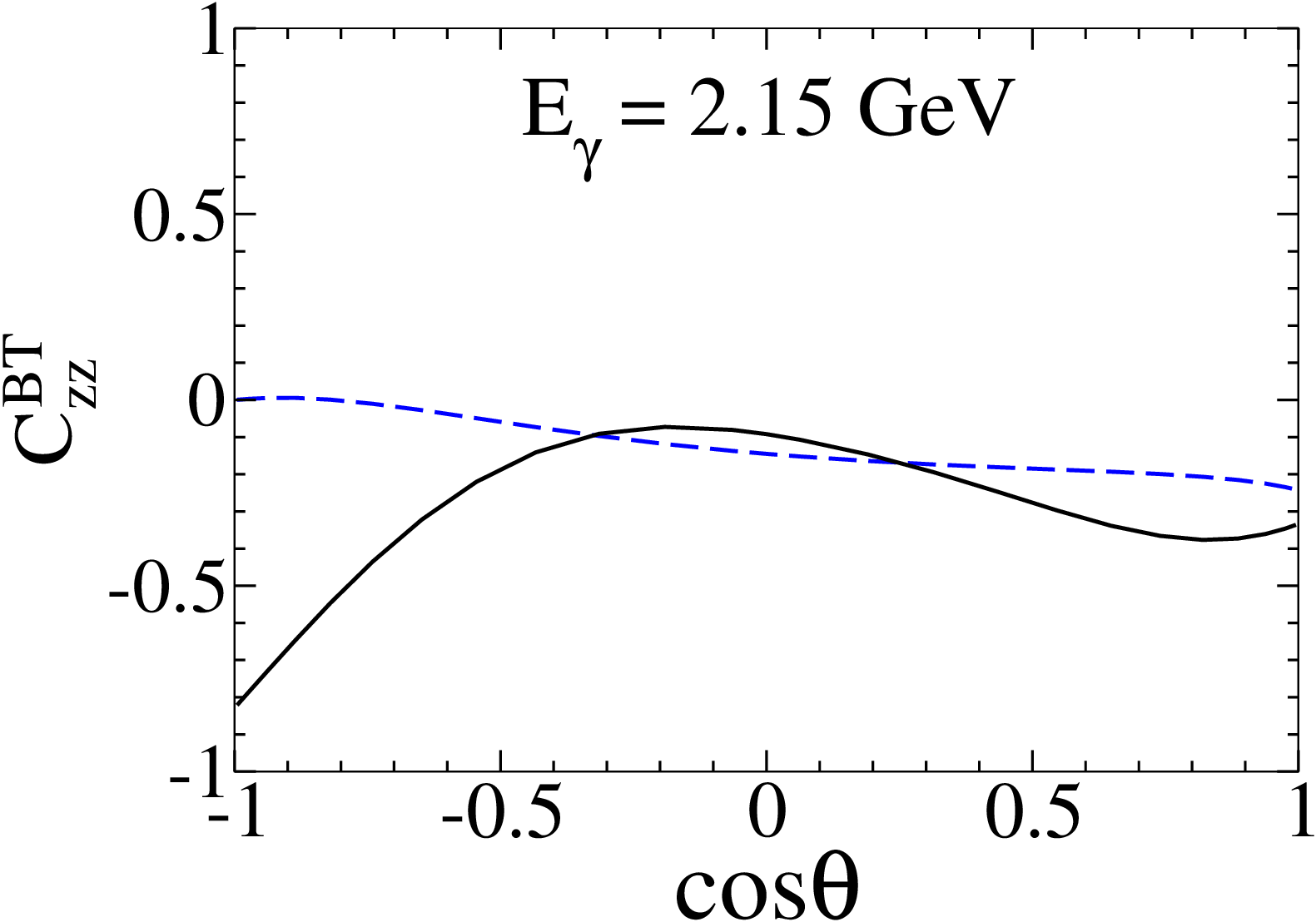}\hspace{3.0em}
\includegraphics[width=6.5cm]{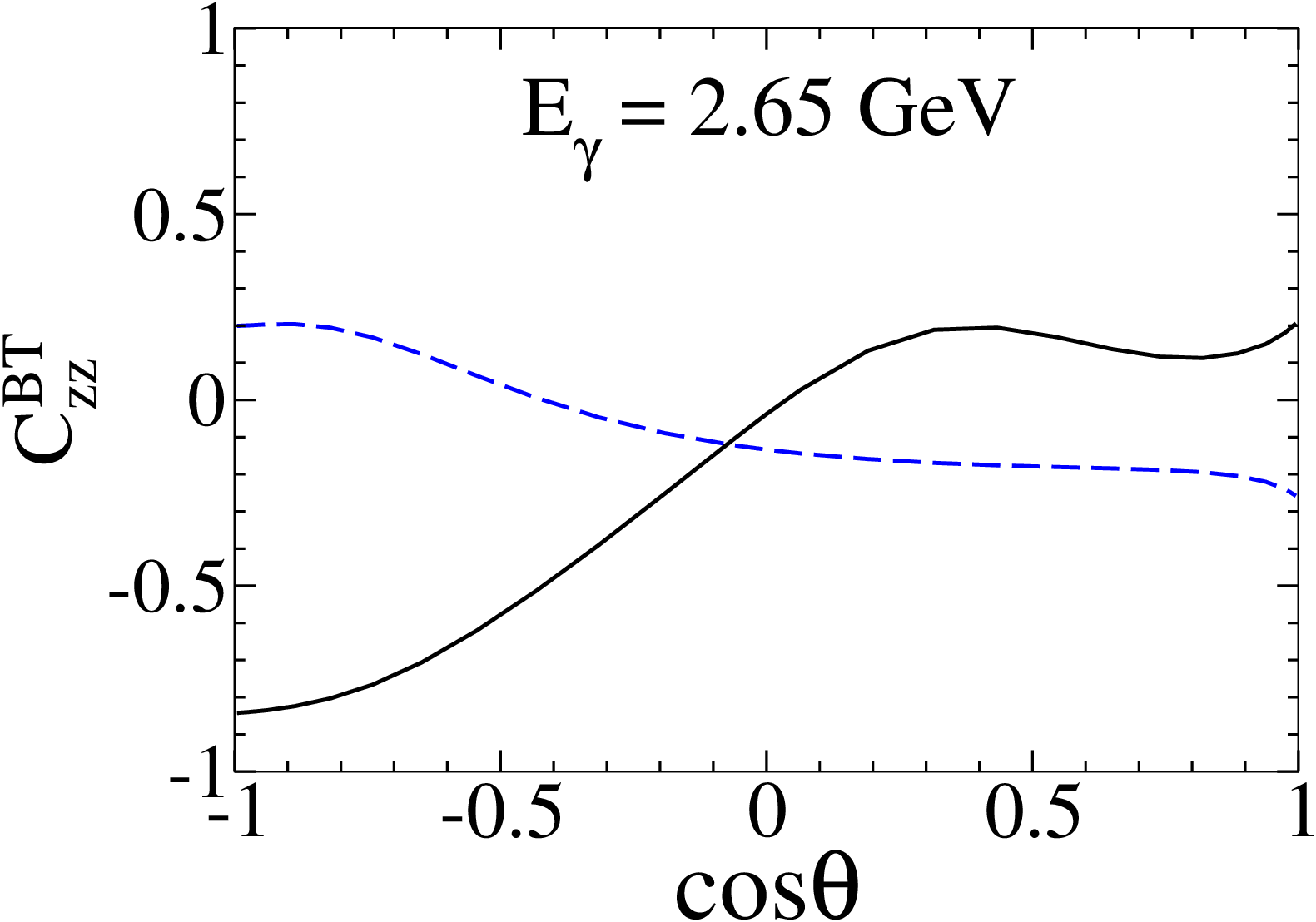}\vspace{1.7em}\\
\includegraphics[width=6.5cm]{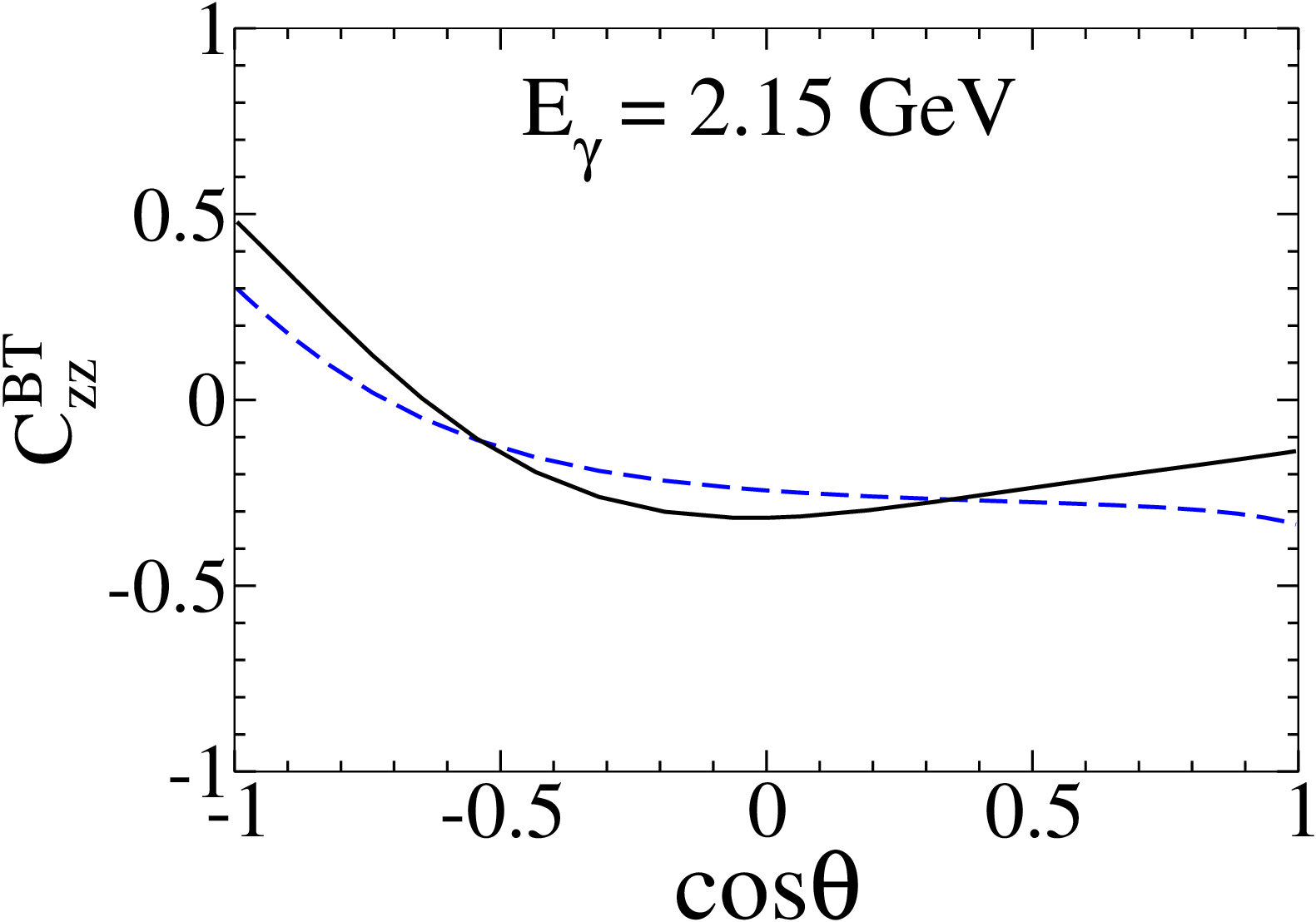}\hspace{3.0em}
\includegraphics[width=6.5cm]{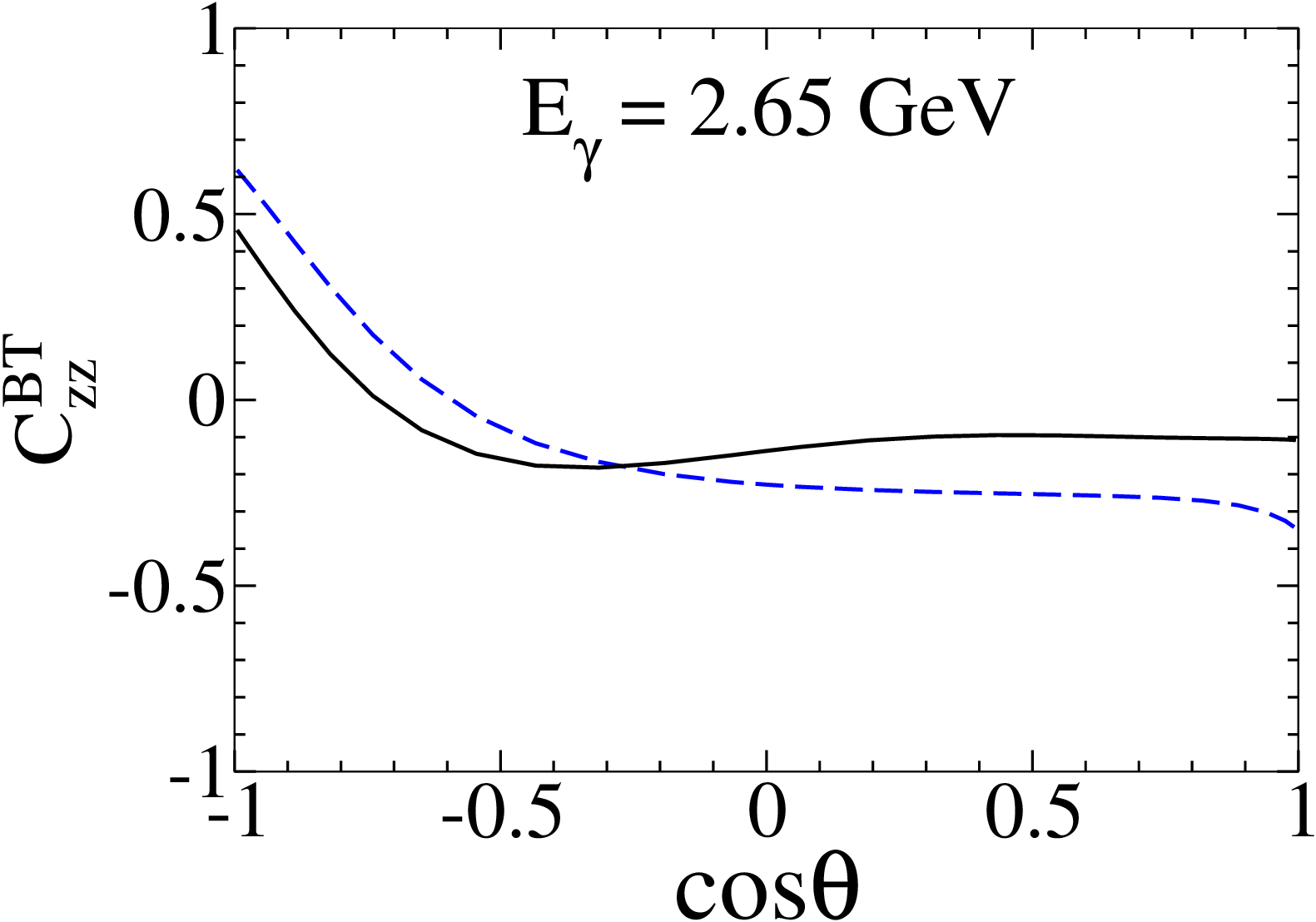}
\end{tabular}
\caption{(Color online) The beam-target asymmetries as functions of 
  $\cos\theta$ with two different photon energies, $E_\gamma=2.15
  \,\mathrm{GeV}$ and $E_\gamma=2.65  \,\mathrm{GeV}$.
In the upper panel, we draw the $C_{zz}^{BT}$ for the $\gamma p\to
K^{*+}\Lambda$ reaction, while in the lower panel we do for the
$\gamma n\to K^{*0}\Lambda$ reaction. 
Notations are the same as in Fig.~\ref{FIG7}.} 
\label{FIG10}
\end{figure}
As drawn in the upper panel of Fig.~\ref{FIG10}, the effects of the
$N^*$ resonances on the BT asymmetry for the $\gamma p \to
K^{*+}\Lambda$ reaction are prominent in comparison to the results
without the $N^*$. While the $C_{zz}^{BT}$ vanishes at the very
backward angle ($\cos\theta=-1$) without the $N^*$ resonances, the
inclusion of them bring its value down to be negative ($\approx
0.8$). It indicates that the polarization of 
the proton highly depends on the $N^*$ resonances. Interestingly, the
effects of the $N^*$ resonances are not at all lessened even at a
higher $E_\gamma$. As $E_\gamma$ increases, the value of the
$C_{zz}^{BT}$ turns positive in the forward angle. The effects of the
$ N^*$ resonances on the neutral process are different from those as
on the charged one, as depicted in the lower panel of
Fig.~\ref{FIG10}. However , in this case, the BT asymmetry becomes 
positive in the very backward direction, and then turns negative as
$\cos\theta$ increases.  

\begin{figure}[ht]
\begin{tabular}{cc}
\includegraphics[width=6.5cm]{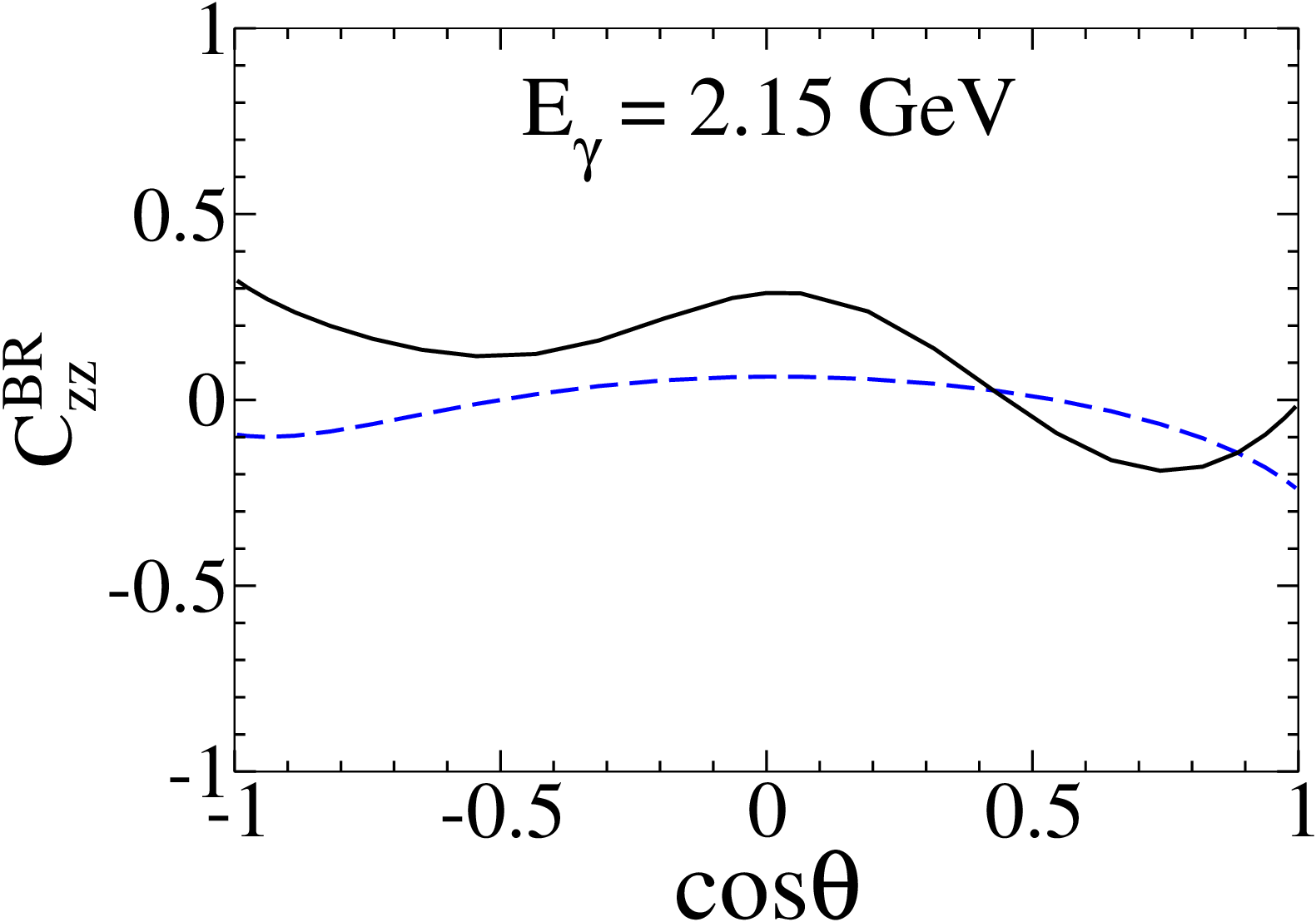}\hspace{3.0em}
\includegraphics[width=6.5cm]{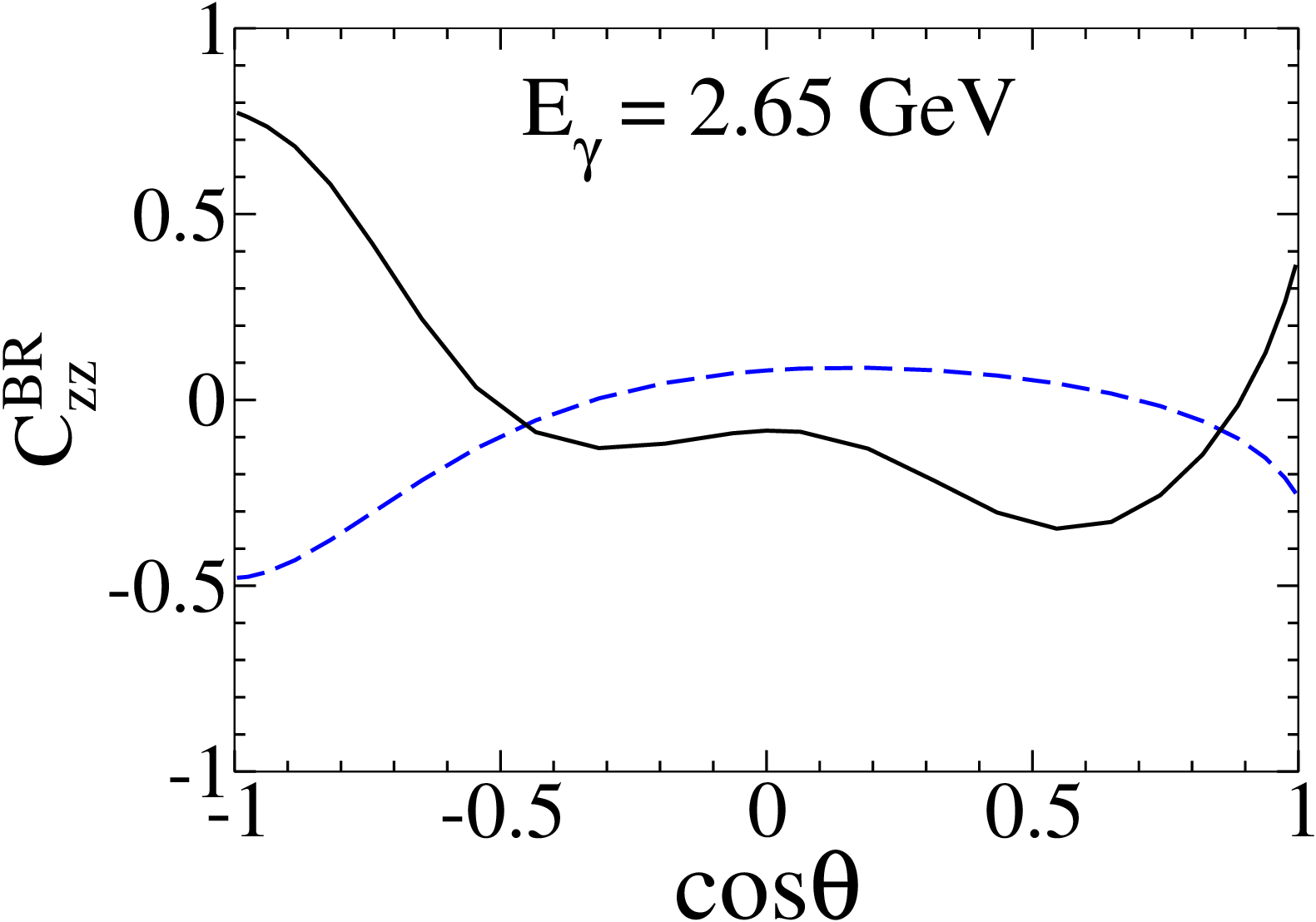}\vspace{1.7em}\\
\includegraphics[width=6.5cm]{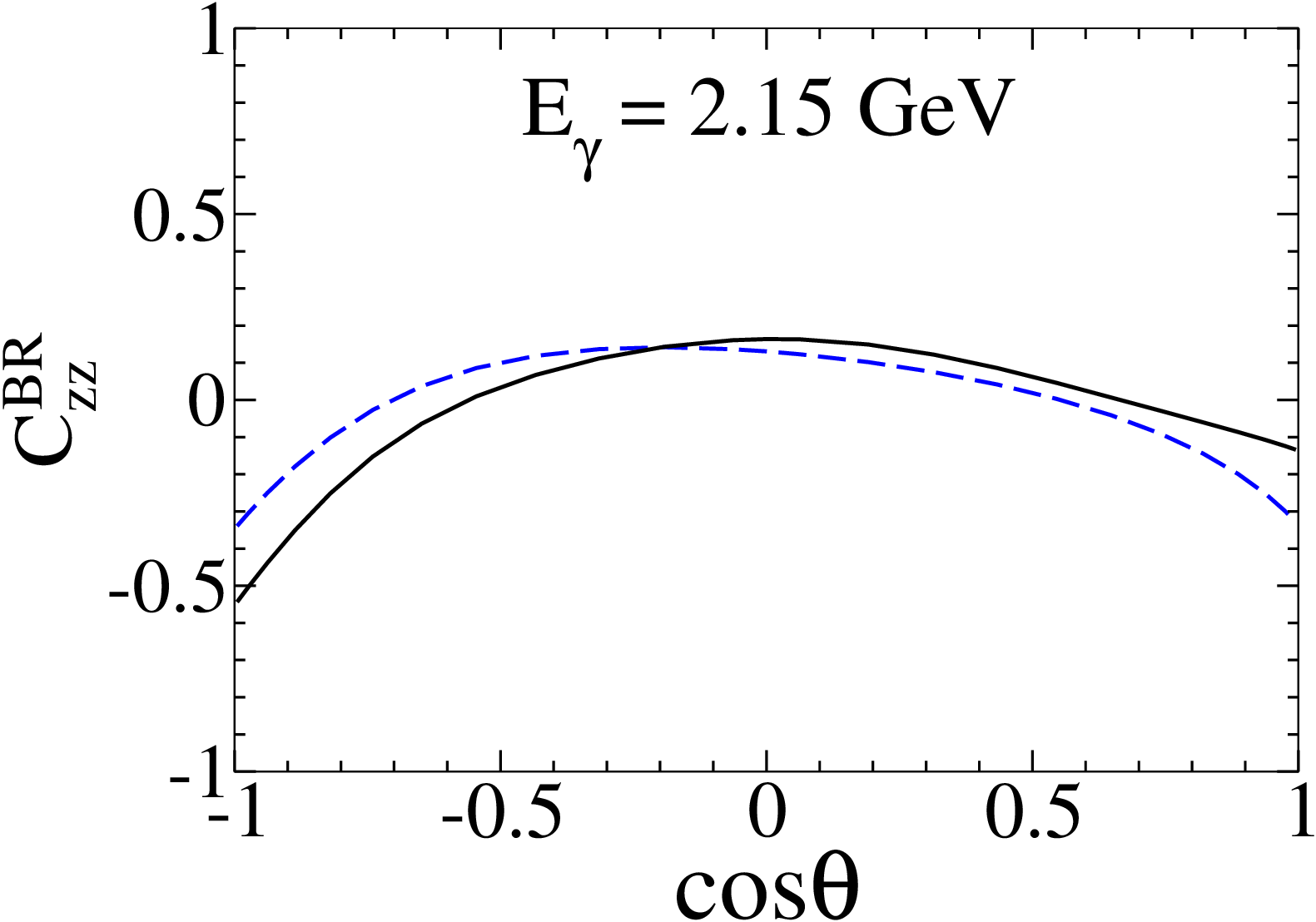}\hspace{3.0em}
\includegraphics[width=6.5cm]{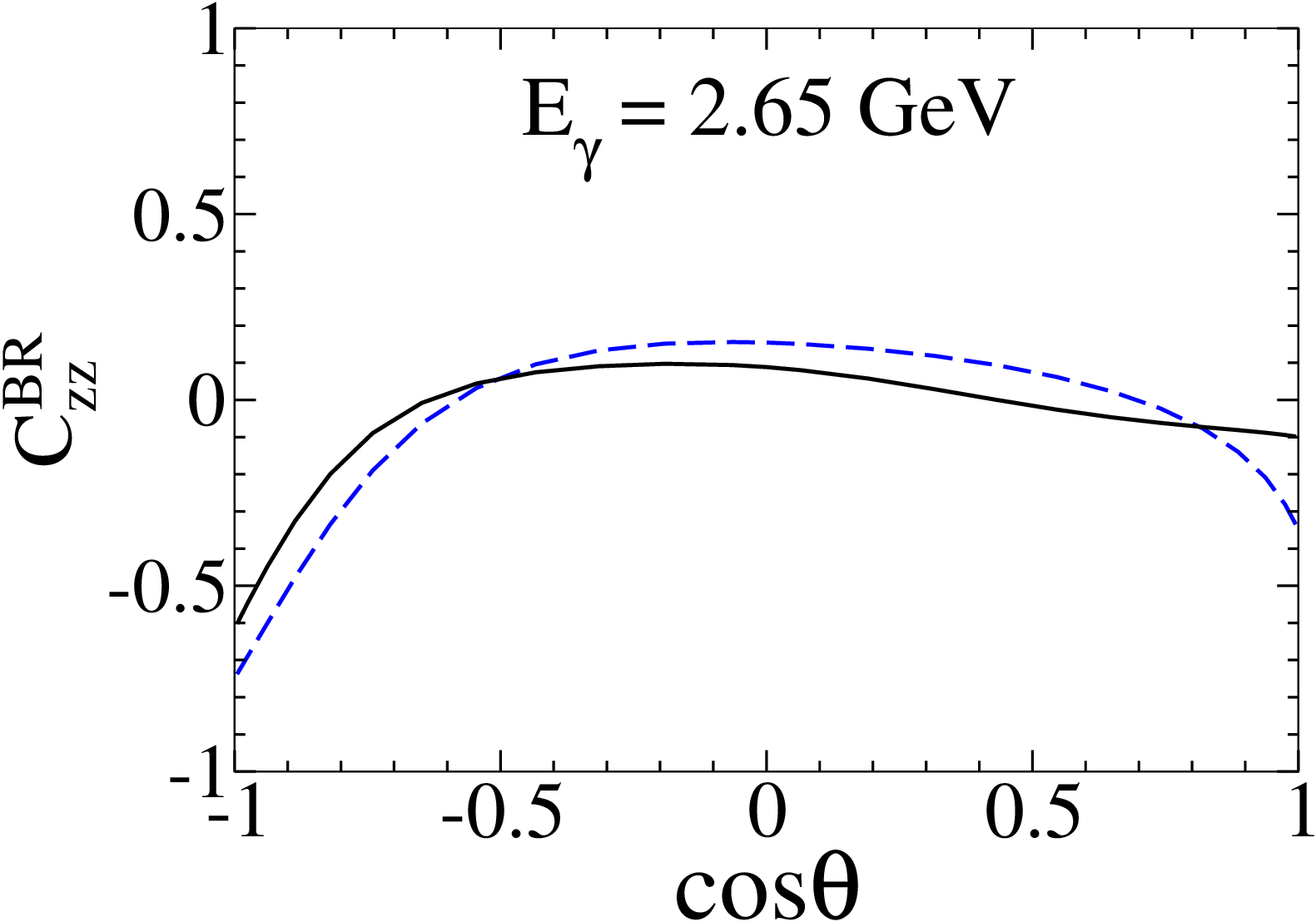}
\end{tabular}
\caption{(Color online) The beam-recoil asymmetries as functions of 
  $\cos\theta$ with two different photon energies, $E_\gamma=2.15
  \,\mathrm{GeV}$ and $E_\gamma=2.65  \,\mathrm{GeV}$.
In the upper panel, we draw the $C_{zz}^{BR}$ for the $\gamma p\to
K^{*+}\Lambda$ reaction, while in the lower panel we do for the
$\gamma n\to K^{*0}\Lambda$ reaction. 
Notations are the same as in Fig.~\ref{FIG7}.} 
\label{FIG11}
\end{figure}
\begin{figure}[ht]
\begin{tabular}{cc}
\includegraphics[width=6.5cm]{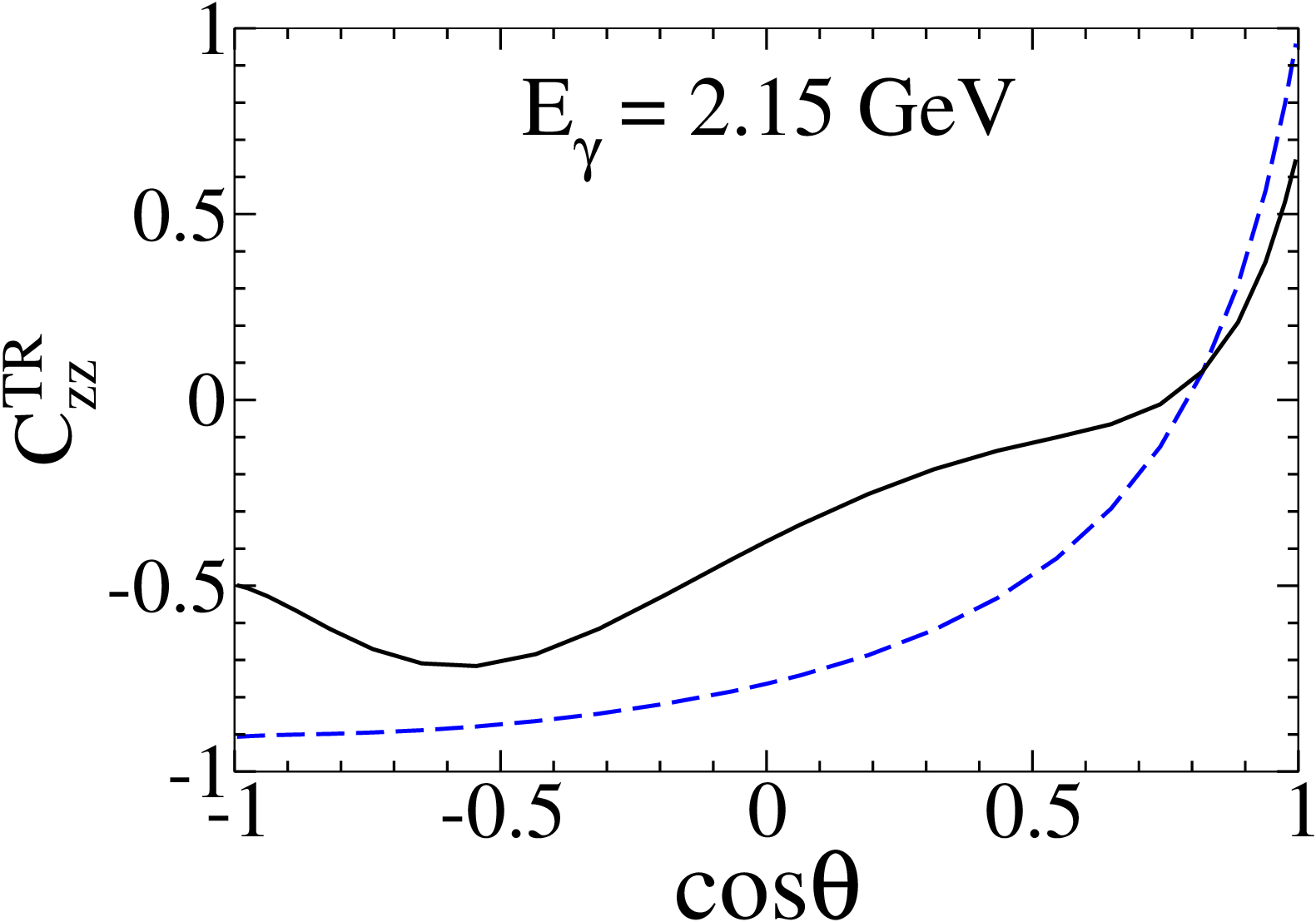}\hspace{3.0em}
\includegraphics[width=6.5cm]{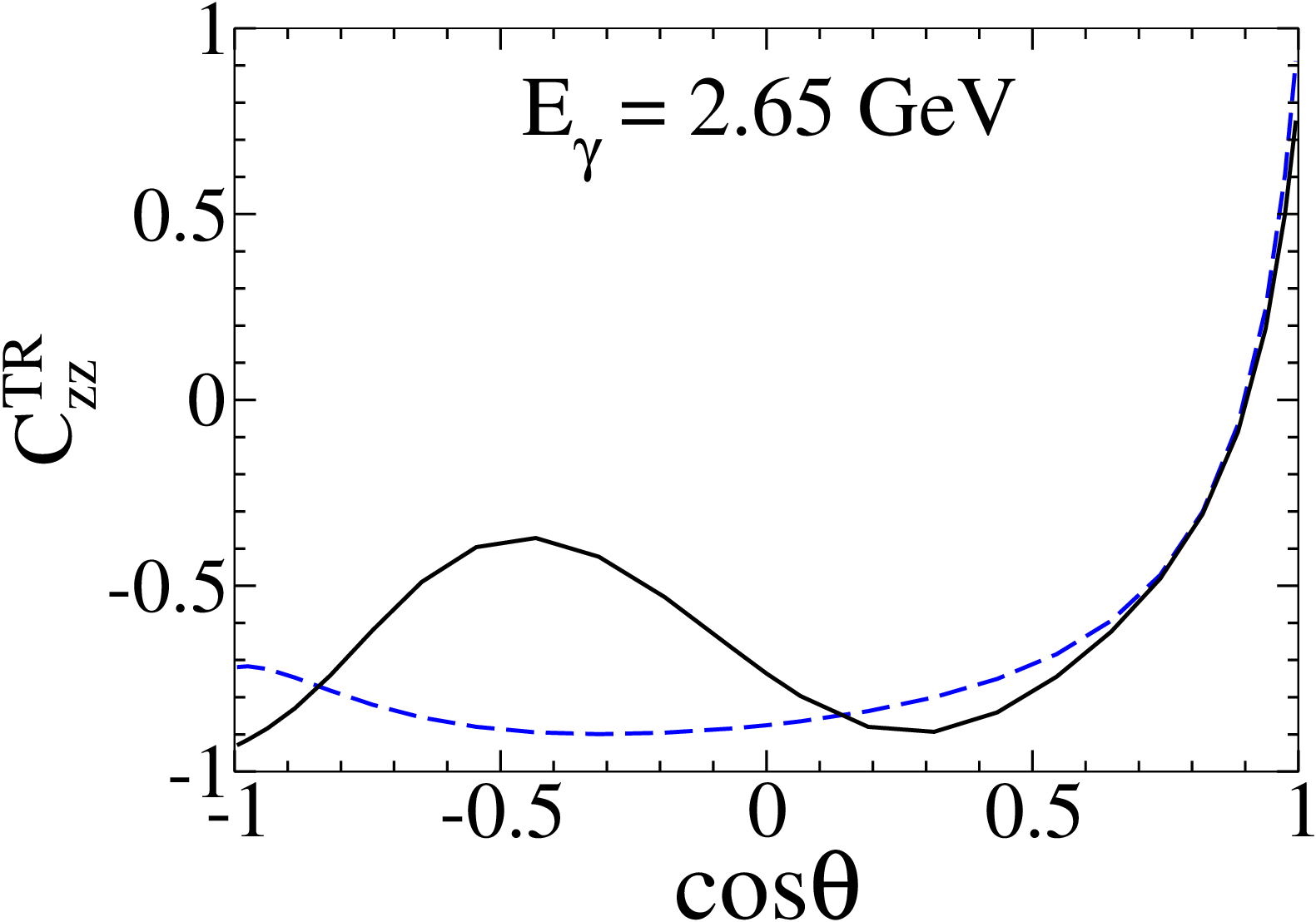}\vspace{1.7em}\\
\includegraphics[width=6.5cm]{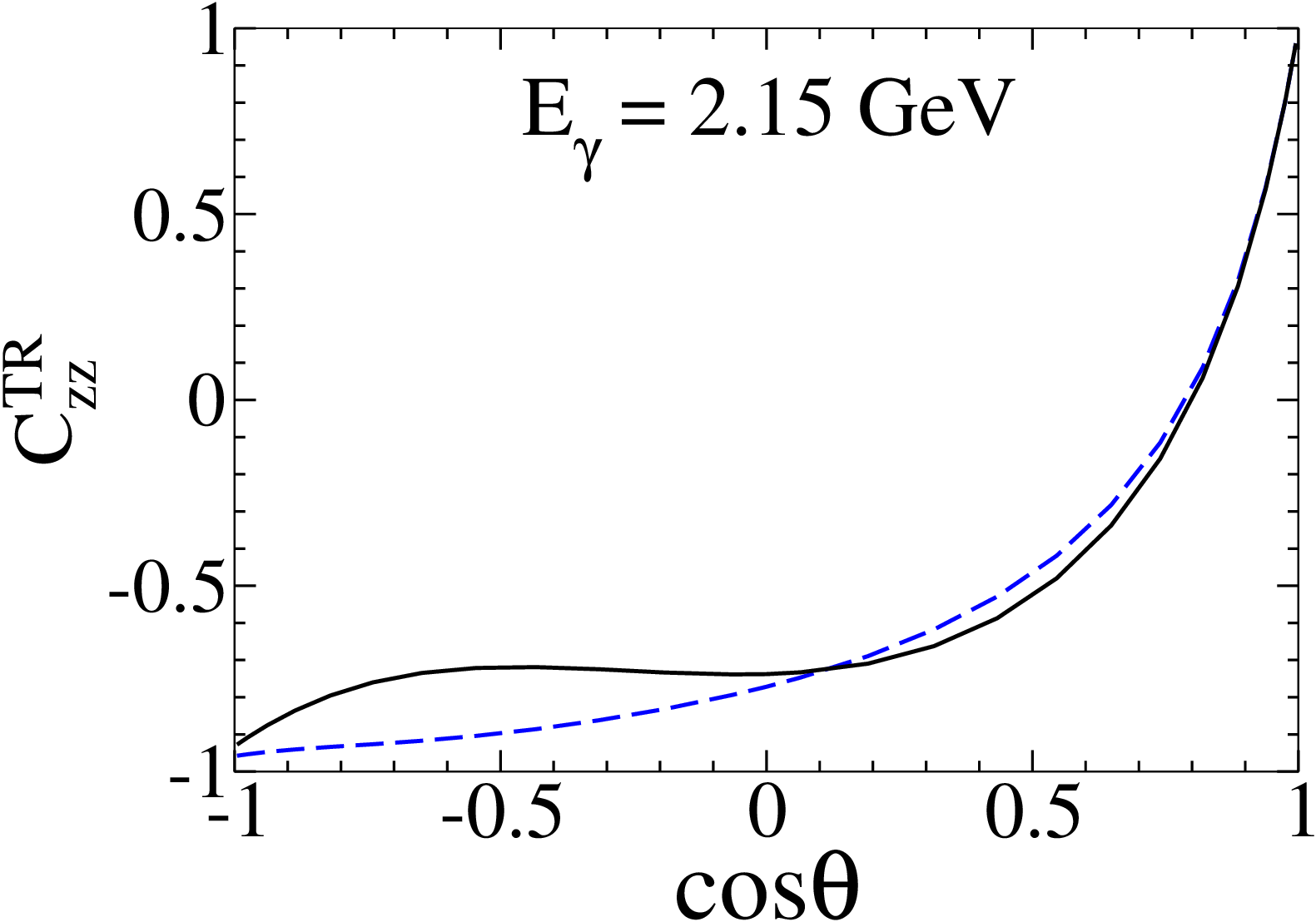}\hspace{3.0em}
\includegraphics[width=6.5cm]{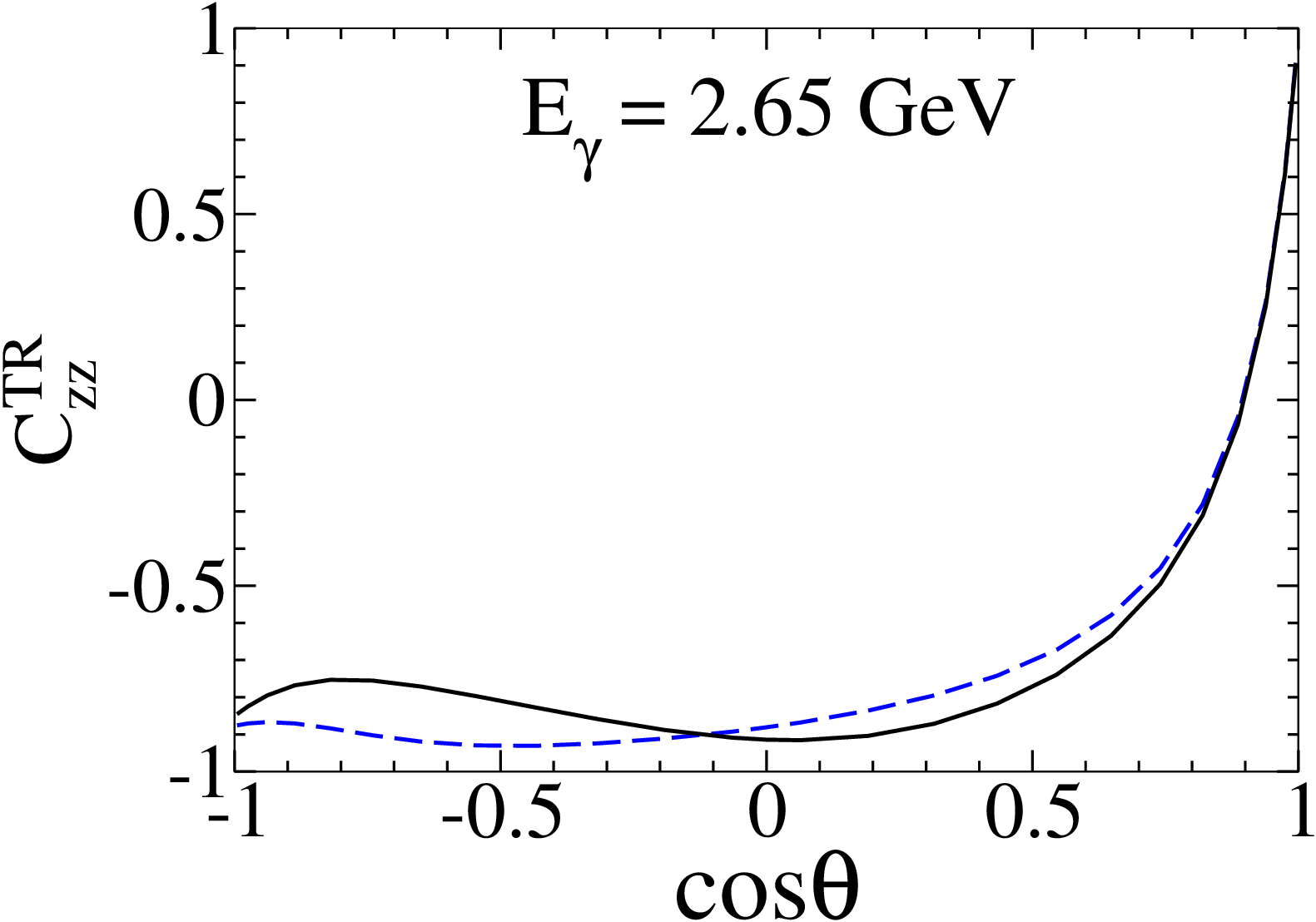}
\end{tabular}
\caption{(Color online) The target-recoil asymmetries as functions of 
  $\cos\theta$ with two different photon energies, $E_\gamma=2.15
  \,\mathrm{GeV}$ and $E_\gamma=2.65  \,\mathrm{GeV}$.
In the upper panel, we draw the $C_{zz}^{TR}$ for the $\gamma p\to
K^{*+}\Lambda$ reaction, while in the lower panel we do for the
$\gamma n\to K^{*0}\Lambda$ reaction. 
Notations are the same as in Fig.~\ref{FIG7}.}
\label{FIG12}
\end{figure}
\begin{figure}[ht]
\begin{tabular}{cc}
\includegraphics[width=6.5cm]{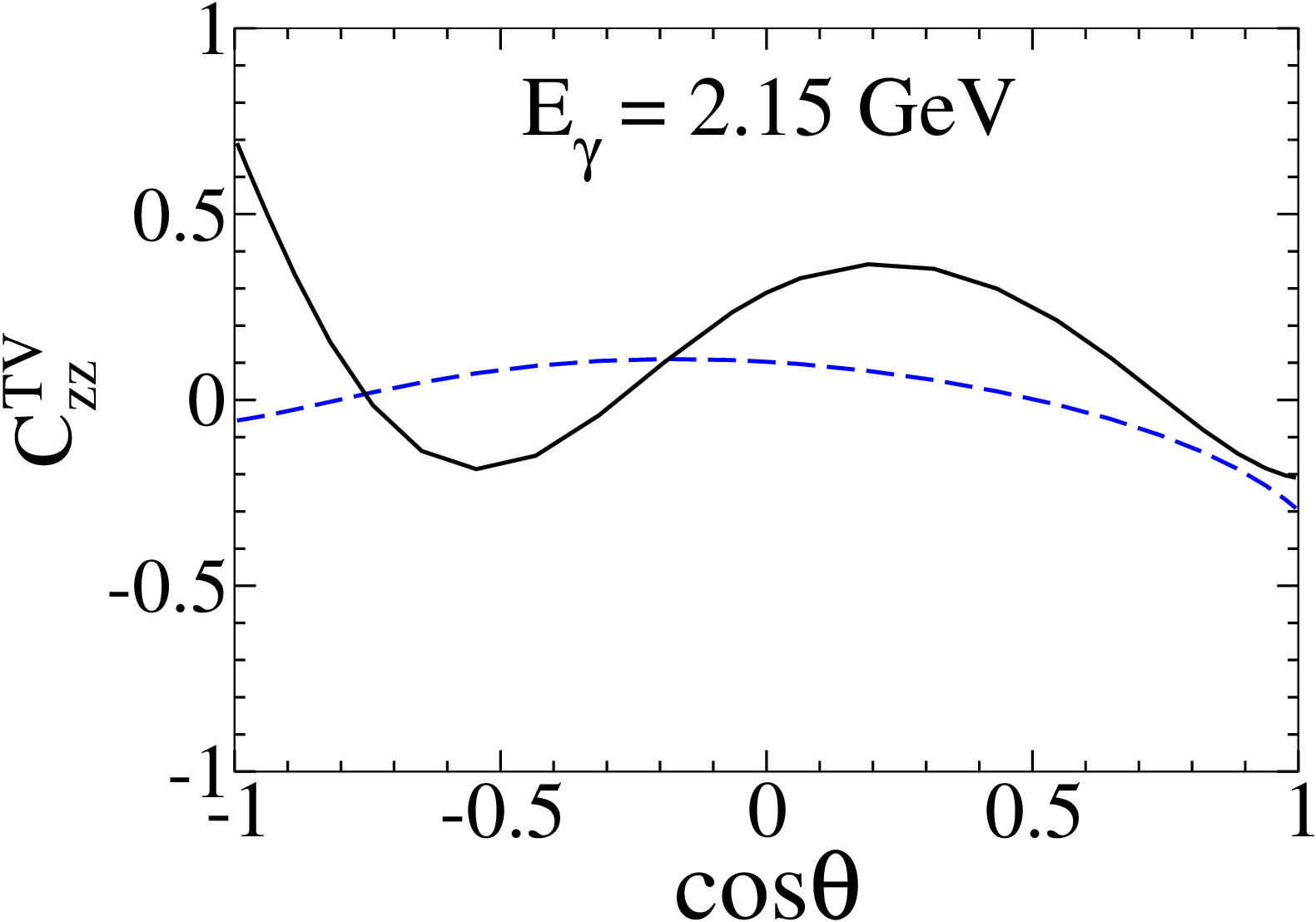}\hspace{3.0em}
\includegraphics[width=6.5cm]{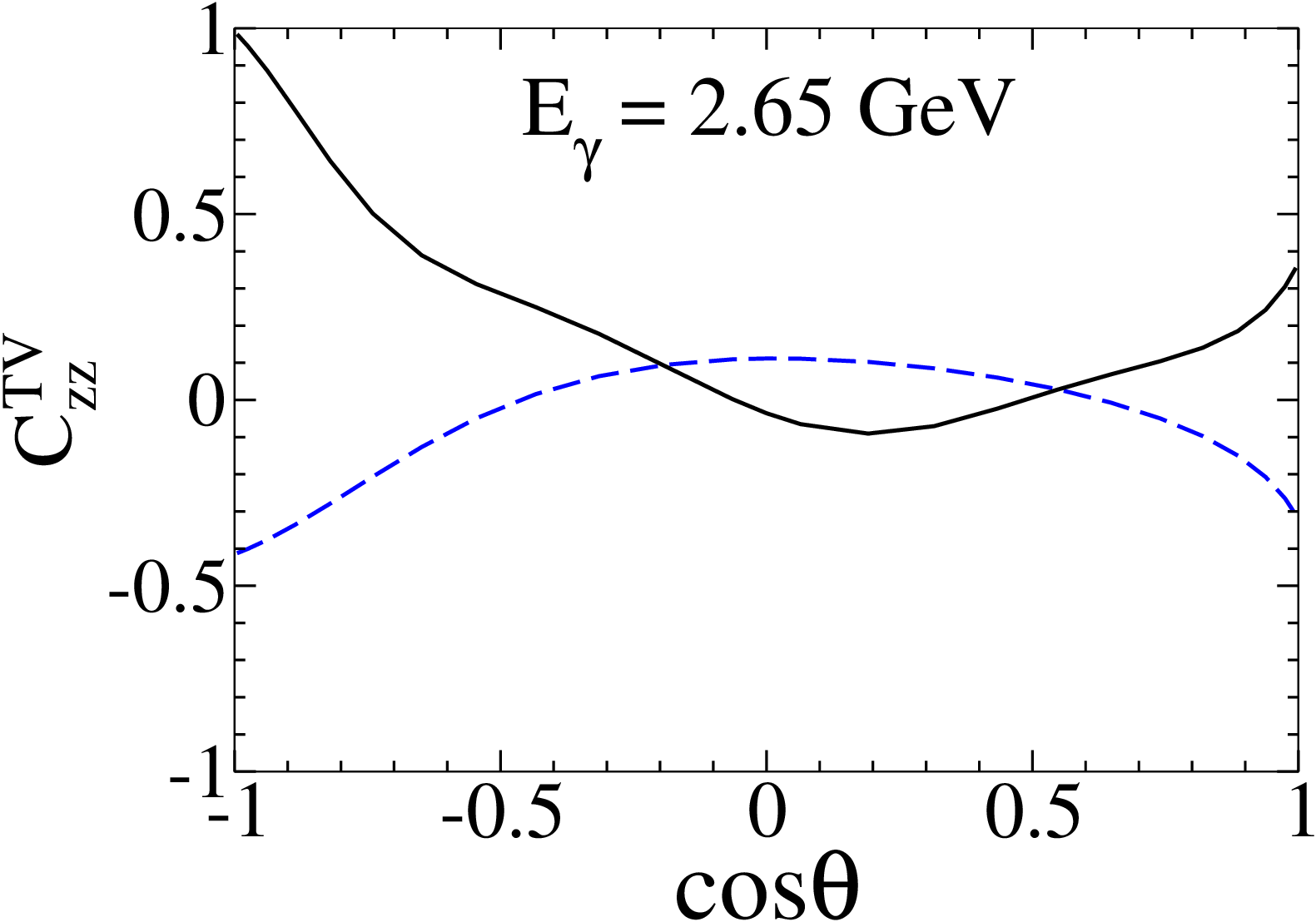}\vspace{1.7em}\\
\includegraphics[width=6.5cm]{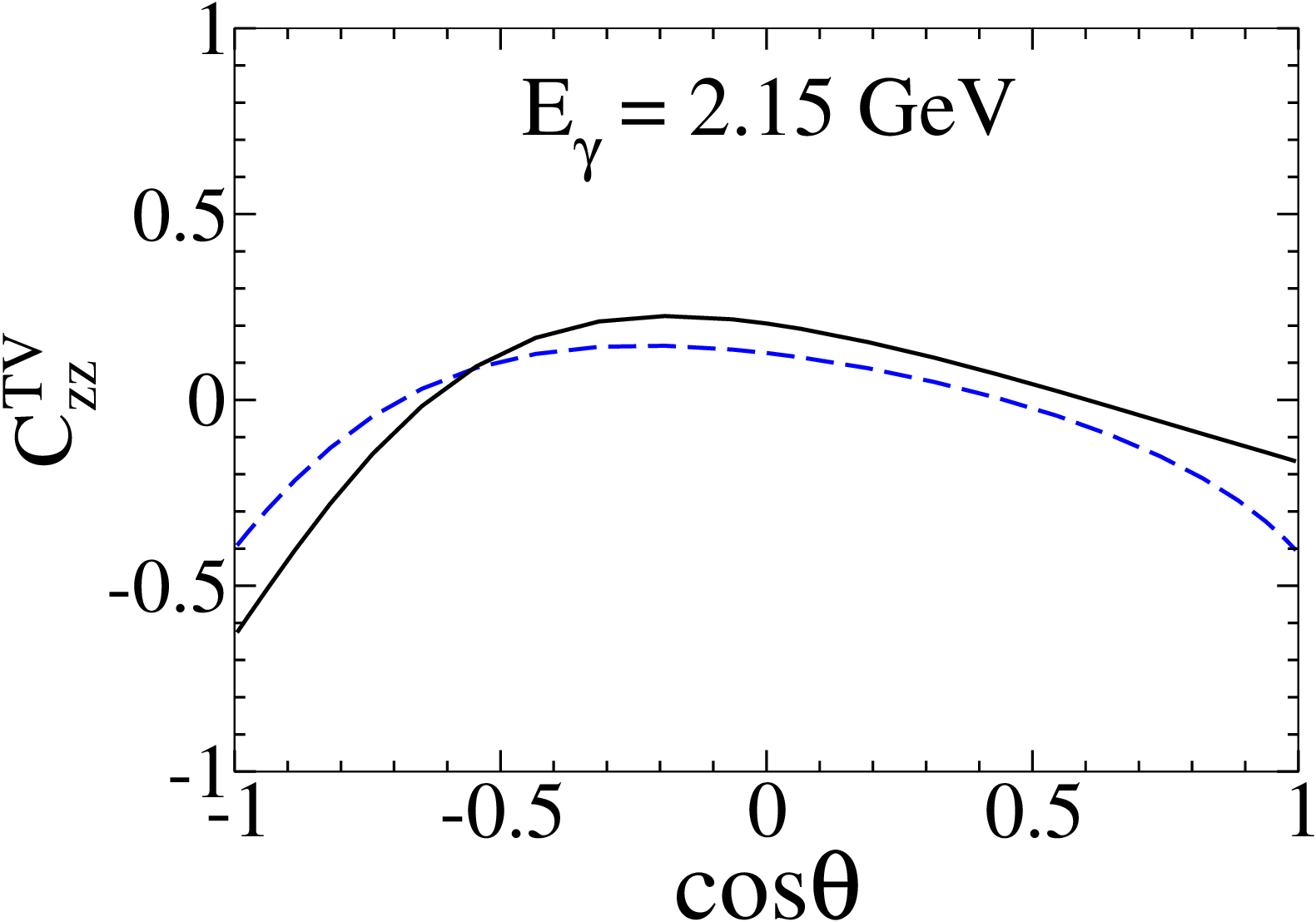}\hspace{3.0em}
\includegraphics[width=6.5cm]{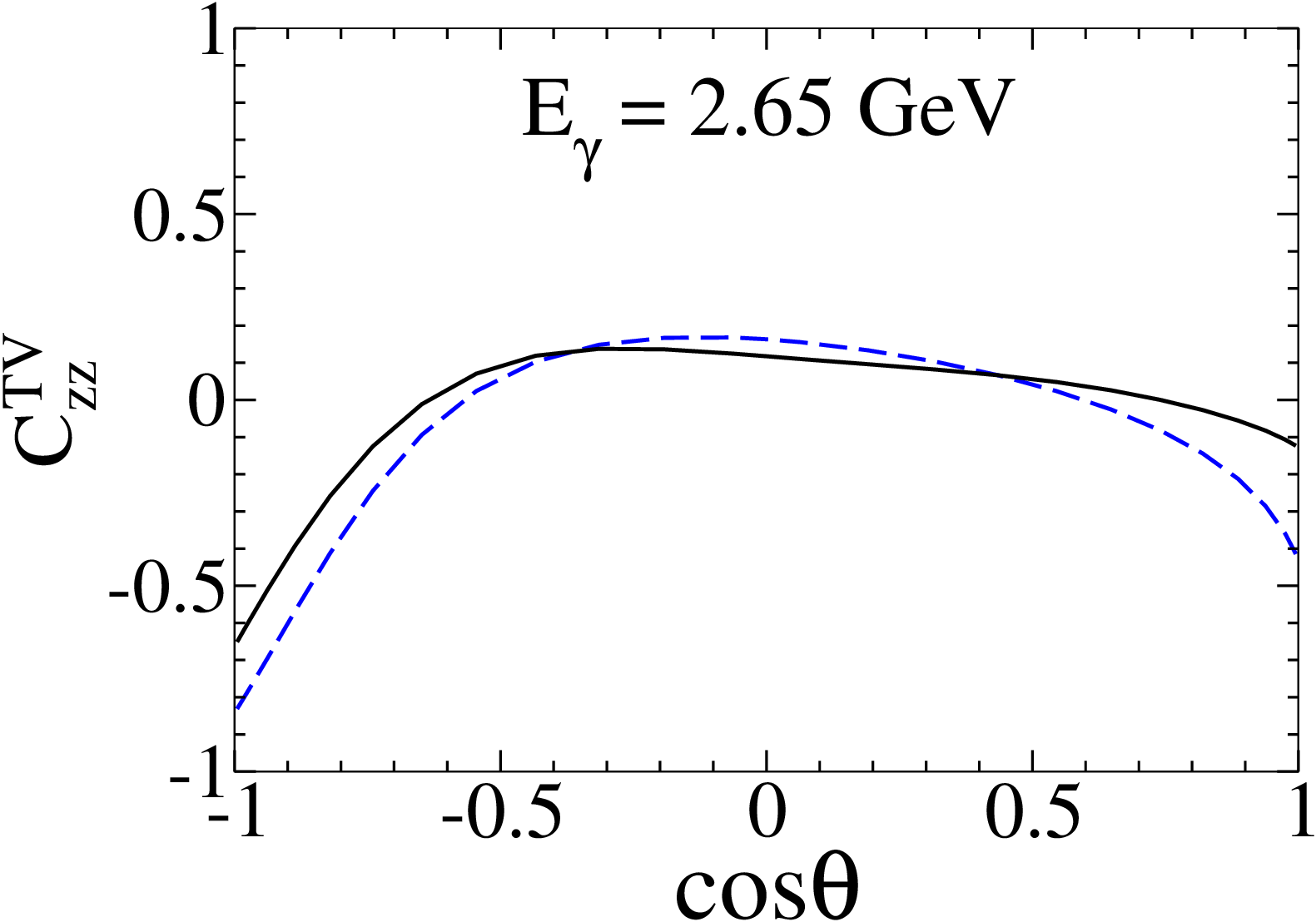}
\end{tabular}
\caption{(Color online) The target-vector-meson asymmetries as
  functions of $\cos\theta$ with two different photon energies,
  $E_\gamma=2.15 \,\mathrm{GeV}$ and $E_\gamma=2.65  \,\mathrm{GeV}$.
In the upper panel, we draw the $C_{zz}^{TV}$ for the $\gamma p\to
K^{*+}\Lambda$ reaction, while in the lower panel we do for the
$\gamma n\to K^{*0}\Lambda$ reaction. 
Notations are the same as in Fig.~\ref{FIG7}.}
\label{FIG13}
\end{figure}
\begin{figure}[ht]
\begin{tabular}{cc}
\includegraphics[width=6.5cm]{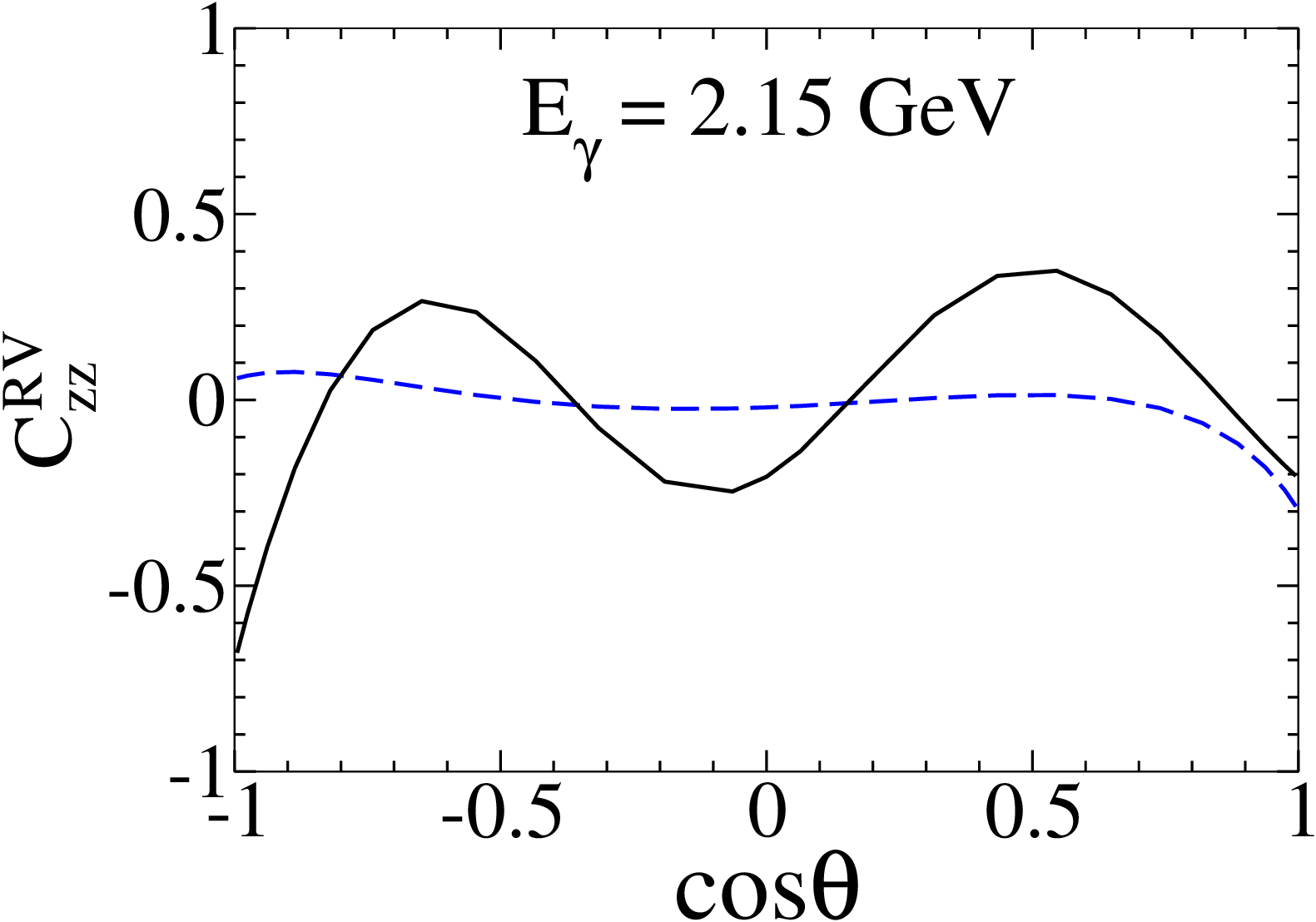}\hspace{3.0em}
\includegraphics[width=6.5cm]{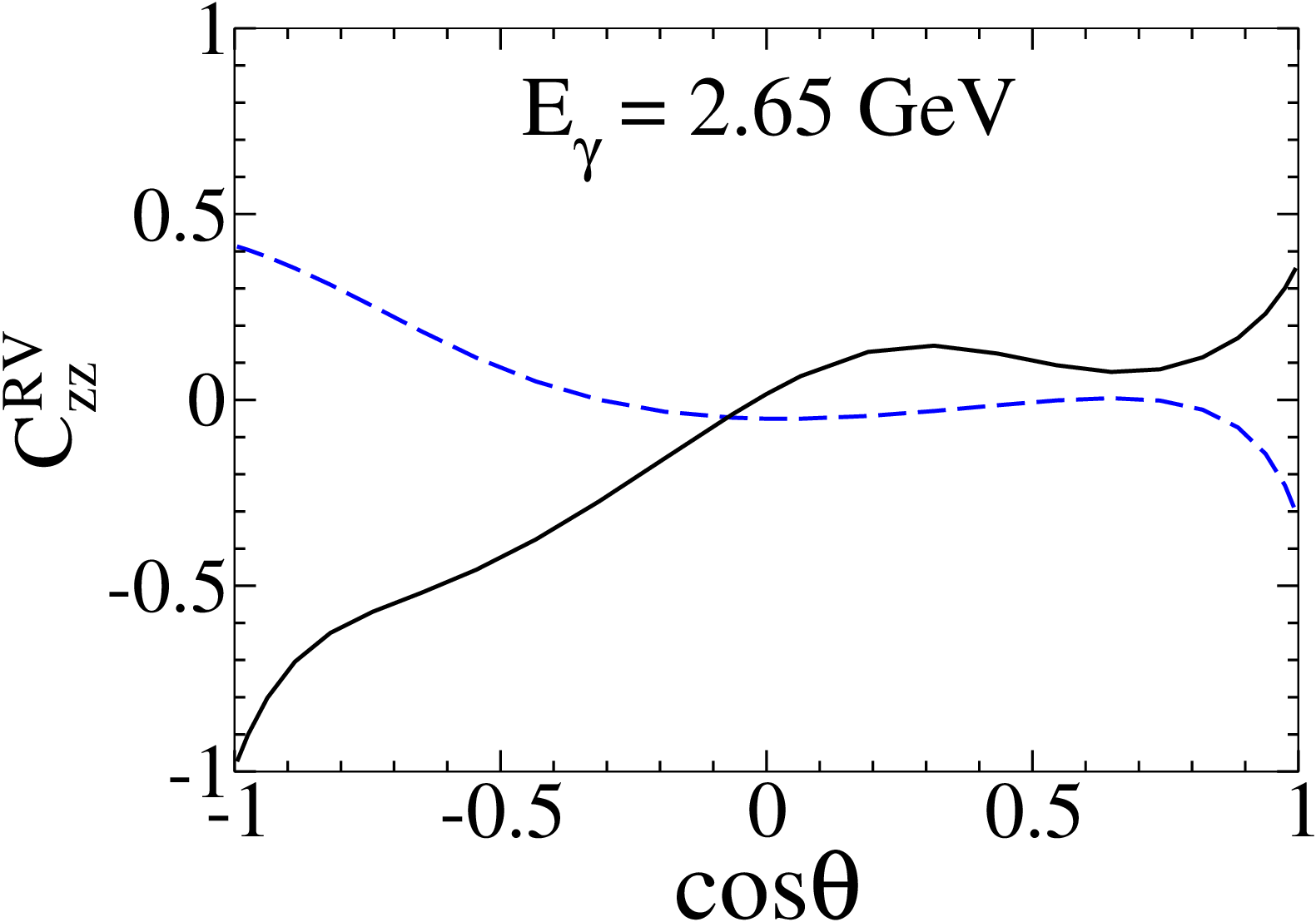}\vspace{1.7em}\\
\includegraphics[width=6.5cm]{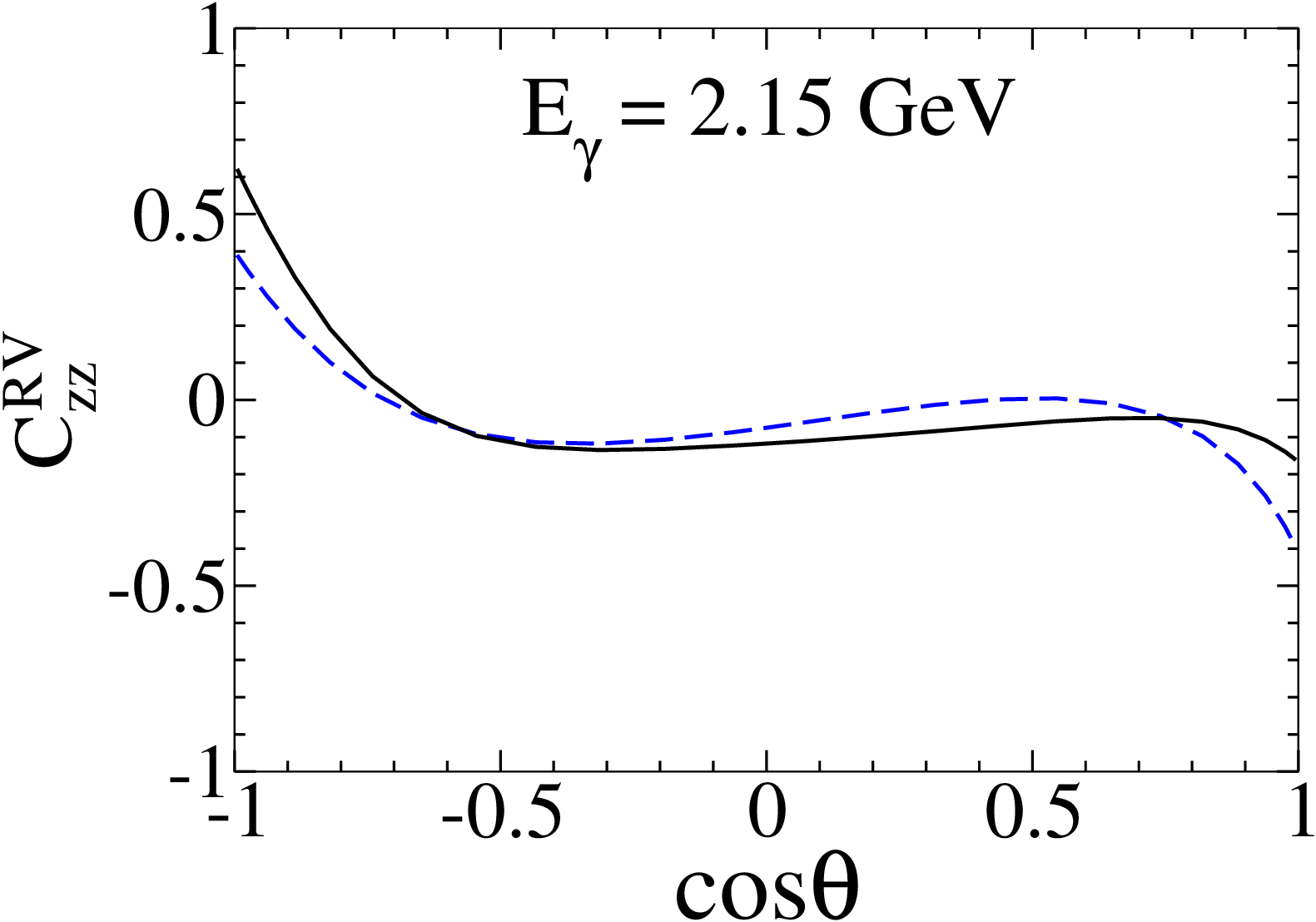}\hspace{3.0em}
\includegraphics[width=6.5cm]{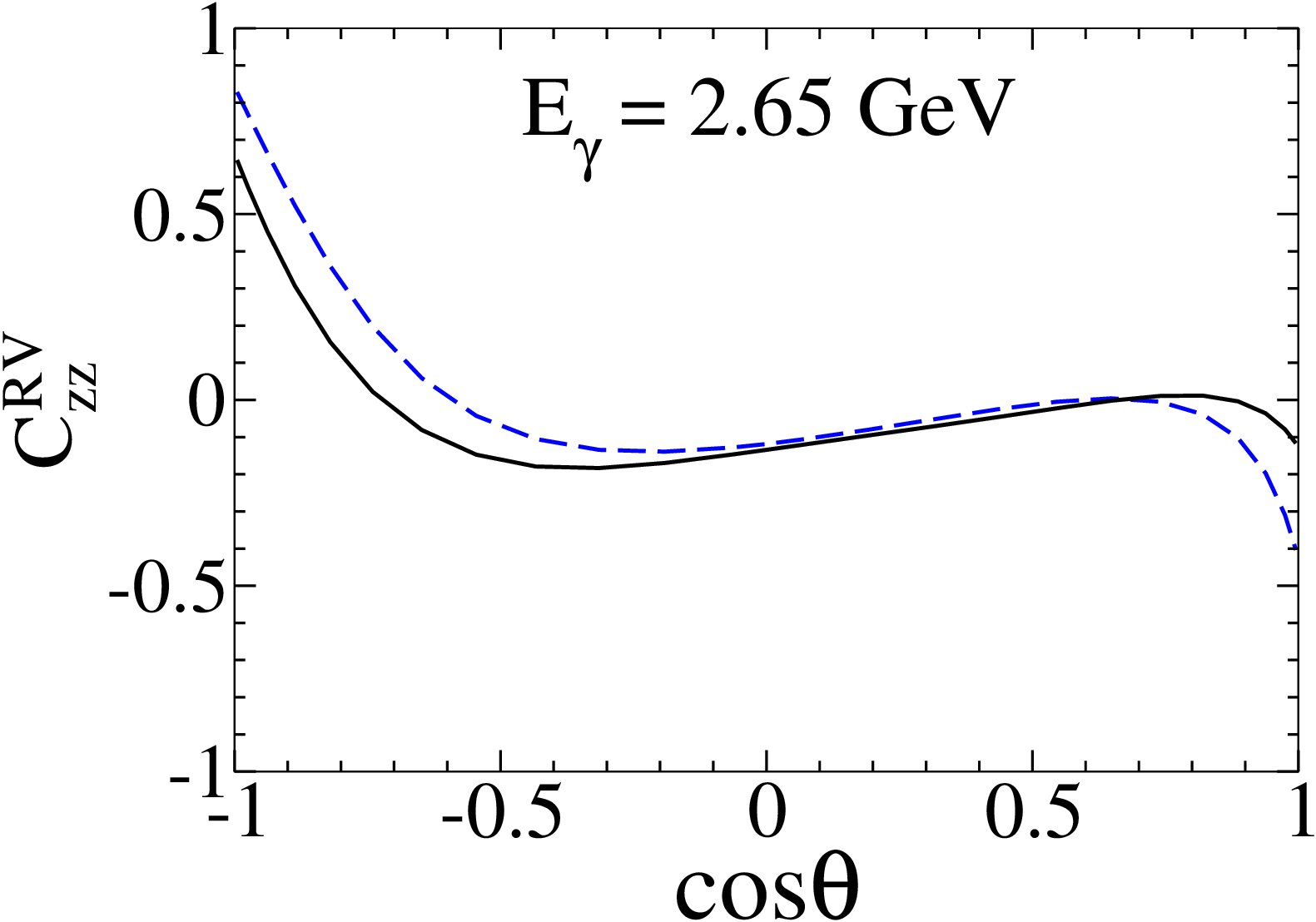}
\end{tabular}
\caption{(Color online) The recoil-vector-meson asymmetries as
  functions of $\cos\theta$ with two different photon energies,
  $E_\gamma=2.15 \,\mathrm{GeV}$ and $E_\gamma=2.65  \,\mathrm{GeV}$.
In the upper panel, we draw the $C_{zz}^{RV}$ for the $\gamma p\to
K^{*+}\Lambda$ reaction, while in the lower panel we do for the
$\gamma n\to K^{*0}\Lambda$ reaction. 
Notations are the same as in Fig.~\ref{FIG7}.}
\label{FIG14}
\end{figure}

The upper and lower panels of Fig.~\ref{FIG11} depict the BR 
asymmetries for the $\gamma p \to K^*\Lambda$ and $\gamma n \to
K^{*0} \Lambda$ reactions, respectively. Again, the effects of the
$N^*$ resonances on $C_{zz}^{BR}$ are clearly seen in the case of
the charged reaction. On the other hand, the $N^*$ effects are
marginal for the neutral channel. We come to the same conclusion for
the TR, TV, and RV asymmetries, as shown in
Fig.~\ref{FIG12}--Fig.~\ref{FIG14}, respectively. Future measurements
of the double polarization observables will be crucial to scrutinizing
the role of the $N^*$ resonances in the $\gamma N\to K^*\Lambda$
reactions.

\section{Summary and conclusion}
In the present work, we aimed at investigating the role of the $N^*$
resonances in explaining the production mechanism of $K^*\Lambda$
photoproduction. We included the following $N^*$ resonances,
$N(2000)\, 5/2^+$, $N(2060)\,5/2^-$, $N(2120)\,3/2^-$, and
$N(2190)\,7/2^-$ in the vicinity of the threshold, based on the PDG
2012 edition. The coupling constants for the electromagnetic and
strong vertices were fixed by the available experimental data
or by theoretical predictions.  
The cut-off masses were determined phenomenologically within a limited
range around 1 GeV. The results of the total cross sections were in 
good agreement with the new CLAS data. In particular,  the two $N^*$ 
resonances $N(2120)\,3/2^-$, and $N(2190)\,7/2^-$ played very
important roles in reproducing the experimental data of the total
cross section for the $\gamma p\to K^{*+} \Lambda$ reaction. The
differential cross sections were also well described in the range of
$1.7\ge E_\gamma \ge 3.9$ GeV, except for the forward angle data in
higher photon energies. We predicted the total and differential cross
sections for the $\gamma n\to K^{*0} \Lambda$ reaction. 
It turned out that the $N^*$ resonances come into play in the
charged channel, whereas their effects on the neutral channel are
marginal. Since the new data for $K^{*0}\Lambda$ photoproduction will
soon be reported~\cite{Mattione}, it will be of great interest to
compare the present results with the upcoming CLAS data. 
We also computed the observables of the single and double
polarizations for the $\gamma N\to K^* \Lambda$ reactions. First, the 
photon-beam asymmetries, the recoil asymmetries, and  the target
asymmetries were studied. The contribution of the $N^*$ resonances
to the single spin asymmetries is prominent in the $\gamma p\to K^{*+}
\Lambda$ reaction, while it is less noticeable for $K^{*0}\Lambda$
photoproduction. The five double polarization observables were computed  
in addition to the single polarization ones, that is, the beam-target
(BT), the beam-recoil (BR), the target-recoil (TR), 
the target-vector-meson (TV), and the
recoil-vectot-meson (RV) asymmetries. We came to the similar
conclusion that the $N^*$ resonances govern the angular dependence
of the double polarization observables, while their effects are in
general marginal for the $\gamma n\to K^{*0} \Lambda$ reaction.

As we discussed in the present work, vector-meson photoproduction
is especially interesting, since its spin structure has a profound
feature coming from the vector meson. We have investigated the 
effects of the $N^*$ resonances at the Born level in an effective
Lagrangian approach, though the $N^*$ resonances turned out to be
essential in describing $K^*\Lambda$ photoproduction, other effects
might be comparably important, in particular, for the spin
observables. For example, the $\gamma N\to K^* \Lambda$ can be
regarded as a subprocess of the $\gamma N\to K\pi \Lambda$ reaction. 
It implies that $K^*\Lambda$ photoproduction may be strongly
coupled to another subprocess such as the $\gamma N\to
K\Sigma^*(1385)$ reaction. Thus, it is also of great interest to
investigate both the $\gamma N\to K^* \Lambda$ and $\gamma N\to
K\Sigma^*(1385)$ processes within a coupled-channel formalism. The
corresponding investigation is under way.  

\section*{Acknowledgments}
We are grateful to K. Hicks, W. Tang, and P. Mattione for the invaluable
discussion about the new CLAS data. H.-Ch. K expresses his gratitude
to P. Navratil and R. Woloshyn for their hospitality during his visit
to TRIUMF, where part of the work was done. S.H.K. is supported by
Scholarship of the Ministry of Education, Culture, Science and
Technology of Japan. A.H. is supported in part by Grant-in-Aid for
Science Research on Priority Areas titled ``Elucidation of New Hadrons
with a Variety of Flavors'' (Grant No. E01:21105006). H.Ch.K is
supported by Basic Science Research Program through the National
Research Foundation of Korea funded by the Ministry of Education,
Science and Technology (Grant Number: 2012001083).

\end{document}